\documentclass[a4paper]{article}

\usepackage[pages=all, color=black, position={current page.south}, placement=bottom, scale=1, opacity=1, vshift=5mm]{background}
\SetBgContents{
}      

\usepackage[margin=1in]{geometry} 

\usepackage{amsmath}
\usepackage{amsthm}
\usepackage{amssymb}

\usepackage[utf8]{inputenc}
\usepackage{hyperref}
\hypersetup{
	unicode,
	pdfauthor={Perugia, G., Paetzel-Prüssman, M., Hupont, I., Varni, G., Chetouani, M., Peters, C., & Castellano, G.},
	pdftitle={Does the Goal Matter?},
	pdfsubject={Does the Goal Matter? Emotion Recognition Tasks Can Change the Social Value of Facial Mimicry towards Artificial Agents},
	pdfkeywords={Human-Robot Interaction, Human-Agent Interaction, Affective Computing, Facial Mimicry, Anthropomorphism, Uncanny Valley, Facial Action Coding System},
	pdfproducer={LaTeX},
    pdfcreator={pdflatex}
}

\usepackage[sort&compress,numbers,square]{natbib}
\theoremstyle{plain}

\theoremstyle{definition}

\newcommand{\cf}{cf.}

\newcommand{\me}{\textit{M}}
\newcommand{\sd}{\textit{SD}}

\usepackage{enumitem}

\usepackage{graphicx, color}
\graphicspath{{fig/}}

\usepackage{algorithm, algpseudocode} 
\usepackage{mathrsfs} 

\usepackage{lipsum}

\title{Does the Goal Matter? Emotion Recognition Tasks Can Change the Social Value of Facial Mimicry towards Artificial Agents}
\author{Giulia Perugia$^{1, *}$, Maike Paetzel-Prüsmann$^{2}$, Isabelle Hupont$^{3,4}$, Giovanna Varni$^{5}$, \\ Mohamed Chetouani$^{4}$, Christopher Peters$^{6}$ and Ginevra Castellano$^{1}$}

\date{
	\texttt{giulia.perugia@it.uu.se}\\

\vspace{0.3cm}$^{1}$ Uppsala Social Robotics Lab, Department of Information Technology, Uppsala University, Uppsala, Sweden \\
$^{2}$ Computational Linguistics, Department of Linguistics, University of Potsdam, Potsdam, Germany \\
$^{3}$ Joint Research Centre, European Commission, Seville, Spain\\
$^{4}$ Institut des Syst\`emes Intelligents et de Robotique, Sorbonne Universit\'e, Paris, CNRS UMR 7222, France \\
$^{5}$LTCI, T\'el\'ecom Paris, Institut Polytechnique de Paris, France \\
$^{6}$KTH Royal Institute of Technology, Department of Computational Science Technology, Stockholm, Sweden \\
}

\begin{document}
	\maketitle
\begin{abstract}
\noindent In this paper, we present a study aimed at understanding whether the embodiment and humanlikeness of an artificial agent can affect people's spontaneous and instructed mimicry of its facial expressions. The study followed a mixed experimental design and revolved around an emotion recognition task. Participants were randomly assigned to one level of humanlikeness (between-subject variable: humanlike, characterlike, or morph facial texture of the artificial agents) and observed the facial expressions displayed by a human (control) and three artificial agents differing in embodiment (within-subject variable: video-recorded robot, physical robot, and virtual agent). To study both spontaneous and instructed facial mimicry, we divided the experimental sessions into two phases. In the first phase, we asked participants to observe and recognize the emotions displayed by the agents. In the second phase, we asked them to look at the agents' facial expressions, replicate their dynamics as closely as possible, and then identify the observed emotions. In both cases, we assessed participants' facial expressions with an automated Action Unit (AU) intensity detector. Contrary to our hypotheses, our results disclose that the agent that was perceived as the least uncanny, and most anthropomorphic, likable, and co-present, was the one spontaneously mimicked the least. Moreover, they show that instructed facial mimicry negatively predicts spontaneous facial mimicry. Further exploratory analyses revealed that spontaneous facial mimicry appeared when participants were less certain of the emotion they recognized. Hence, we postulate that an emotion recognition goal can flip the social value of facial mimicry as it transforms a likable artificial agent into a distractor. Further work is needed to corroborate this hypothesis. Nevertheless, our findings shed light on the functioning of human-agent and human-robot mimicry in emotion recognition tasks and help us to unravel the relationship between facial mimicry, liking, and rapport.
\end{abstract}

\noindent\textbf{Keywords}: Human-Robot Interaction, Human-Agent Interaction, Affective Computing, Facial Mimicry, Anthropomorphism, Uncanny Valley, Facial Action Coding System

\maketitle

\section{Introduction}

The success of artificial agents in areas like healthcare, personal assistance, and education highly depends on whether people perceive them as likable and pleasant to interact with. In the lab, people's perceptions of an artificial agent can be easily measured with questionnaires and interviews. In real-life settings, instead, the artificial agent is on its own and the explicit evaluation of the interaction is not always feasible. In fact, in these contexts, people might skip the proposed surveys or reply carelessly due to lack of time and interest \cite{chung2018your}.
A more promising approach in such contexts may be the use of behavioral measures. While behavioral measures are in general extensively used in Human-Agent and Human-Robot Interaction (HAI and HRI), they are seldomly linked to people's self-reported perceptions (e.g., likability and engagement, see \cite{perugia2020engage, perugia2021ICanSeeIt}). In this paper, we focus on facial mimicry - the mirroring of another person’s facial expressions \cite{hatfield1992primitive} - as a target behavioral cue. In particular, we are interested in understanding \textit{whether humans mimic the facial expressions of the six basic emotions displayed by artificial agents, and how this can be linked to their perceptions of the agents}.

From psychology, we know that facial mimicry increases with rapport \cite{tickle1990nature, hess1995intensity}, but also appears in first acquaintances between individuals as a sign of liking \cite{chartrand1999chameleon, kulesza2015face}.
Studies on facial mimicry in HAI and HRI have so far mostly focused on whether artificial agents are liked better when given the ability to mimic a human interaction partner \cite{hoegen2018impact, numata2020achieving, riek2009anthropomorphism}. Hofree et al. \cite{hofree2014bridging} were among the only researchers who investigated whether human interaction partners mimic the facial expressions of artificial agents as well. In their study, they disclosed that people's mimicry of an android's facial expressions of anger and happiness is connected with their perceptions of the agent's humanlikeness only when the android is co-present. In our study, we extend Hofree et al.'s work \cite{hofree2014bridging} by (i) including a \textit{wider spectrum of artificial agents}, (ii) employing an overall \textit{less realistic humanoid robot that allows for easy alteration of facial cues} (i.e., Furhat), (iii) focusing on \textit{all six basic emotions}, and (iv) using a \textit{computer vision technique} in lieu of Electromyography (EMG) to estimate people's facial mimicry. With respect to EMG, computer vision is far less obtrusive and hence more viable for field use.

In this study, we involve 45 participants in an emotion recognition task with three artificial agents varying in embodiment (i.e., physical Furhat robot, video-recorded Furhat robot, and a virtual agent) and humanlikeness (i.e., humanlike, characterlike, and morph). 
The emotion recognition task used in our experiment was divided into two phases. In the first phase, participants were asked to observe the facial expressions of the six basic emotions as expressed by the three artificial agents and a video-recorded human (i.e., the control), and pick the correct one from a list. In the second phase, instead, they were asked to observe the same facial expressions in re-shuffled order, mimic their temporal dynamics as closely as possible, and only afterwards recognize them. Based on Kulesza et al. \cite{kulesza2015face}, this latter phase was carried out under the pretense that intentional mimicry of facial expressions could actually improve participants' emotion recognition. Participants' faces were video-recorded in both stages of the experiment and the activation of the action units (AU) corresponding to the six basic emotions was determined through Hupont and Chetouani's AU intensity detector \cite{hupont2019}. In the first part of the study, we gauged which facial expressions were \textit{spontaneously} mimicked by participants. In the second part of the study, we focused instead on participants' \textit{instructed} mimicry, and estimated how accurate participants were in replicating the temporal dynamics of the observed facial expressions.

The aim of this study is to understand (1) whether an artificial agent’s embodiment and humanlikeness can influence people’s spontaneous and instructed facial mimicry (as suggested by Hofree et al. \cite{hofree2014bridging} and Mattheij et al. \cite{mattheij2013vocal, mattheij2015mirror}), (2) if spontaneous facial mimicry is related to people’s perceptions of artificial agents, especially in terms of anthropomorphism, social presence, likability, and uncanniness (perceptual dimensions expected to be influenced by the agent's level of humanlikeness), and (3) whether there is a link between instructed and spontaneous facial mimicry. 
The overarching ambition of this work is to explore \textit{whether spontaneous facial mimicry can be used as an implicit, unconscious cue of liking and rapport in HAI and HRI, and whether instructed facial mimicry can act as its proxy in settings where spontaneous facial mimicry is difficult to gauge}. Our work contributes to efforts paving the way towards unobtrusive automatic assessment of facial mimicry in interactions with artificial agents, hence facilitating the measurement of liking and rapport through behavioral cues in the future.

\section{Related Work}

Facial mimicry is the spontaneous imitation of another individual's facial expression without explicit instruction to do so \cite{hatfield1992primitive}. Within the area of facial mimicry research, emotional mimicry refers to the spontaneous mirroring of a facial expression with inherent emotional meaning, for instance, wincing when observing others in pain \cite{bavelas1986show} or frowning at another person's frown. 
This paper focuses on people's mimicry of the six basic emotions - happiness, sadness, surprise, anger, fear, and disgust \cite{ekman1997face} - as displayed through the facial expressions of artificial and human agents. Within the subsections \ref{sec:naturemimicry}, \ref{sec:evidence}, and \ref{sec:socialvalue}, we give an account of the different theories on the nature and functioning of spontaneous facial mimicry in human-human interactions (HHI). We then describe the literature on human-agent and human-robot facial mimicry in subsection \ref{sec:mimicryartificial}, and explain our interest in instructed facial mimicry in subsection \ref{sec:instructedmimicry}.

\subsection{Nature of Spontaneous Facial Mimicry}
\label{sec:naturemimicry}

There are two main theoretical perspectives on the nature and functioning of emotional mimicry: a motor and an emotional perspective. The \textit{motor perspective} holds that emotional mimicry is an unconscious, unintentional, unemotional, and reflex-like matching of observed facial expressions \cite{chartrand1999chameleon}. Within this context, the \textit{associative sequence learning} (ASL) approach posits that mimicry happens by virtue of a learned long-term association between an action stimulus (e.g., a person's smile) and an action response (e.g., the observer's smile; \cite{heyes2011automatic}), which holds as long as the action stimulus (e.g., the observed facial expression) is similar to other stimuli previously associated with a certain motor action (e.g., the observer's facial expression).

Another theoretical formalization within the motor perspective is the \textit{automatic embodiment account}, which postulates that mimicry is the embodied motor simulation of an observed emotion that serves the purpose of emotion recognition \cite{niedenthal2010simulation}. According to this approach, we mimic another individual's facial expressions to better recognize and differentiate them.

As opposed to the motor perspective, the \textit{emotional perspective} sees mimicry as a marker of subtle affective states arising in response to emotional stimuli \cite{dimberg1990distinguished, dimberg1997facial}. Within this perspective, the \textit{facial-feedback hypothesis} \cite{tomkins1984affect, izard2013human}, which dates back to Darwin \cite{darwin1998expression}, posits that ``the sight of a face that is happy, loving, angry, sad, or fearful (...) can cause the viewer to mimic elements of that face and, consequently, to catch the other's emotions'' \cite{hatfield1992primitive}. With a slightly different line of thought, the \textit{affect-matching account} suggests that observing a facial expression triggers a corresponding affective state in the observer, which \textit{then} generates the mimicking act \cite{dimberg2000unconscious}. 
Within the emotional perspective, there is hence no clear consensus yet as to whether the affective state arising from an emotional stimulus precedes or succeeds mimicry.

The motor and emotional perspectives make somewhat different claims on the outcomes of emotional mimicry \cite{moody2007more}. The motor perspective assumes that facial mimicry is always consistent with the observed facial expression (i.e., emotion-congruent mimicry). For instance, an expression of anger can only trigger a corresponding expression of anger. On the opposite, the emotional perspective suggests that mimicry is related to the action tendencies associated with a stimulus (e.g., competitive and collaborative tasks, \cite{lanzetta1989expectations}). Thus an expression of anger can trigger anger but also fear (i.e., valence-congruent mimicry), and the type of emotion triggered depends on the meaning associated with the observed facial expression and the context where mimicry takes place \cite{fischer2012emotional}. 

\subsection{Evidence Supporting Theoretical Accounts on Spontaneous Facial Mimicry}
\label{sec:evidence}

In general, there is little experimental support for the motor perspective. Available studies almost exclusively focused on facial mimicry of happiness and anger. As Hess and Fischer and Hess et al. \cite{hess2013, hess2014} underline, such studies only confirm that people display a valence-congruent facial expression when exposed to happiness and anger (i.e., smiling to happiness, frowning to anger). However, they do not fully back up emotion-congruent facial mimicry, which is at the core of the motor perspective.
With regards to the automatic embodiment account, several studies have investigated whether blocking facial mimicry impairs the correct recognition of emotional facial expressions \cite{niedenthal2001did, hawk2012face}. Current evidence supports this position only partially. Indeed, mimicry seems to be crucial for emotion recognition but only when it comes to recognizing ambiguous or subtle facial expressions \cite{hess2001facial, fischer2012emotional}.

There are a number of studies that support the emotional perspective. For instance, Laird and Bresler \cite{laird1992process} noticed that when people are asked to reproduce facial expressions of fear, anger, sadness, and disgust, they also report experiencing those emotions. Moreover, Ekman et al. \cite{ekman1983autonomic} note that the muscular reproduction of the facial expressions of the six basic emotions activates the Autonomic Nervous System (ANS) in a similar way as to when people actually experience those emotions. Finally, Dimberg et al. \cite{dimberg1998rapid} describe how the facial response system that is responsible for mimicry responds to emotions faster (300-400 ms) than the ANS (1-3 sec.), thus finding support for the affect-matching account. Further support for the emotional perspective was also brought by Moody et al. \cite{moody2007more} who found that fear priming elicits expressions of fear in response to both fear and anger, thus demonstrating that mimicry is not a purely automatic mirroring of an observed emotion, but has an intrinsic emotional meaning.

\subsection{The Social Value of Spontaneous Facial Mimicry}
\label{sec:socialvalue}

Regardless of their different views on the nature of facial mimicry, both the motor and the emotional perspective posit that facial mimicry serves a social purpose. In one case (i.e., motor perspective), it serves to recognize and respond to other people's emotions. In the other case (i.e., emotional perspective), it serves the purpose of emotional contagion \cite{hatfield1993emotional, varni2017computational}, as to say ``the tendency to automatically mimic and synchronize movements, expressions, postures, and vocalizations with those of another person and, consequently, to converge emotionally'' \cite{hatfield1992primitive}.
The literature suggests that mimicry is indicative of higher liking during first acquaintances \cite{chartrand1999chameleon, kulesza2015face, calvo2020effects}, stronger rapport in already established relationships \cite{hess1995intensity} and that it increases when two interaction partners are given the goal to affiliate \cite{lakin2003chameleon}. In fact, \cite{hess1995intensity} found that watching funny movies with friends elicits more laughs than watching them with strangers. Consistently, \cite{fischer2012emotional} discovered that dyads of friends mimic each other's smiles of pride more than strangers do. \cite{hess2013} and \cite{bourgeois2008impact} propose that mimicry acts as a \textit{social regulator} as it communicates the intention to bond. Since emotional mimicry is known to be related with interpersonal stance \cite{prepin2013beyond}, social tuning \cite{bernieri1988coordinated}, bonding \cite{jaques2016understanding}, and rapport \cite{tickle1990nature, gratch2006virtual, wang2009rapport}, we consider it an important phenomenon to study in Human-Robot (HRI) and Human-Agent Interaction (HAI). In fact, if facial mimicry was found to work similarly for artificial agents and humans, it could be used as an implicit and unconscious measure of the quality of interaction in HAI and HRI \cite{perugia2020role}.

\subsection{Spontaneous Facial Mimicry of Virtual Agents and Social Robots}
\label{sec:mimicryartificial}

In face-to-face interactions between humans, acted facial expressions constitute the only possibility of studying spontaneous facial mimicry in a controlled way. However, acted facial expressions can be perceived by humans as being inauthentic and hence might hinder the occurrence of mimicry. For this reason, in psychology, studies on spontaneous facial mimicry have almost exclusively focused on static images or videos of facial expressions, with these latter being sometimes used to simulate live video-sessions \cite{kulesza2015face}. With respect to humans, virtual and robotic agents give the unique possibility to investigate spontaneous mimicry in face-to-face interactions occurring in real-time while preserving control over the experimental setup \cite{hoegen2018impact}. This is because they enable researchers to manipulate only a few facial action units (AU) and control their activation over time. In this sense, the use of virtual and robotic agents not only allows to investigate whether spontaneous facial mimicry occurs or not in specific contexts, but also opens up the possibility to understand whether its temporal dynamics are replicated.

While human-agent mimicry has been explored more thoroughly \cite{gratch2006virtual, hoegen2018impact}, studies on human-robot mimicry gained popularity more recently. Such a delay is probably due to the fact that robots' faces were not provided with enough degrees of freedom to accurately reproduce facial expressions until very recently. Most available studies on human-robot and human-agent mimicry focus on endowing agents with the ability to mimic the facial expressions of human interactants and observing how this ability affects people's perceptions and reactions \cite{hoegen2018impact, numata2020achieving, riek2009anthropomorphism}. Only a few studies investigate people's spontaneous mimicry of an artificial agent's facial expressions.
Such studies show similar results to human-human mimicry, with the main difference residing in the lower intensity and slower speed of human-agent and human-robot mimicry. For instance, Mattheij et al. \cite{mattheij2015mirror, mattheij2013vocal} found evidence for the spontaneous mimicry of happiness, surprise, and disgust in the context of HAI and Philip et al. \cite{philip2018rapid} disclosed that people spontaneously mimic virtual agents' facial expressions of joy, anger, and sadness. They also observed that mimicry is less intense when it is directed to a virtual agent with respect to a human one. Similarly, in HRI, Hofree et al. \cite{hofree2014bridging} observed that people mimic a video-recorded android (i.e., Hanson's Einstein robot) to a lesser extent than a video-recorded human. Furthermore, they discovered that, while the facial expressions of a video-recorded android are mimicked only when the robot is perceived as highly humanlike, physically co-present androids are mimicked regardless of the perceptions they elicit. Hence, they proposed that it is the robot's co-presence that makes its humanlike appearance highly salient, and in turn elicits spontaneous facial mimicry. 
Following this line of thought, in the present study, \textit{we manipulated the artificial agents’ humanlikeness, as well as their embodiment, and attempted to understand whether these influenced spontaneous facial mimicry.}
We employed all three embodiments used by Hofree et al. \cite{hofree2014bridging} - a video-recorded human, a video-recorded robot, and a physical robot. Moreover, we added a virtual agent as in Mattheij et al. \cite{mattheij2013vocal, mattheij2015mirror}. In line with Li \cite{li2015benefit}, we considered: (1) the video-recorded robot as \textit{artificial, physically embodied}, but \textit{not co-present}; (2) the physical robot as \textit{artificial, physically embodied}, and \textit{co-present}; and (3) the video-recorded human as \textit{natural, physically embodied}, but \textit{not co-present}. 
While Li \cite{li2015benefit} differentiates between \textit{physical} and \textit{digital co-presence}, in this work we combined the two into one single category of \textit{co-presence} to distinguish between the two video-recordings that capture behavior of the past and hence do not share the same environment and time with the participant (i.e., video-recorded robot and video-recorded human) from the virtual agent which shares the same environment and time with the participant. Consequently, we categorize the virtual agent as \textit{artificial, virtually embodied}, and \textit{co-present}.

In HHI, Bourgeois and Hess \cite{bourgeois2008impact} showed that the social context in which the interaction takes place has the power to influence emotional mimicry. While happy expressions are mimicked regardless of whether an observed person is an in-group or out-group member, expressions of sadness are mimicked only between in-group members. Likewise, in HRI, Hofree et al. \cite{hofree2018behind} showed that participants mimicked a robot's smiles and frowns when cooperating with it, but displayed inverse mimicry (i.e., frowned at the robot's smiles and smiled in response to its frowns), when the context was competitive.
To circumvent this problem, in this study, we showed the agents' facial expressions to participants in a non-interactive context inspired by Kulesza et al. \cite{kulesza2015face}. Similar to Hofree et al. \cite{hofree2014bridging}, in this study, \textit{we asked participants to carefully observe the agents' facial expressions}. Inspired by Kulesza et al. \cite{kulesza2015face}, however, \textit{we also gave them the goal to recognize the emotion displayed by the agent}.

\subsection{Spontaneous and Instructed Facial Mimicry}
\label{sec:instructedmimicry}

Facial mimicry can further be divided into spontaneous and instructed. Spontaneous facial mimicry, which we have discussed so far, occurs unconsciously, without any specific instruction \cite{hatfield1992primitive}. Instructed facial mimicry, instead, is deliberate mimicry of facial expressions that occurs consciously as a result of specific instructions \cite{mcintosh2006social, paetzel2017investigating}. In their study, Hofree et al. \cite{hofree2014bridging} used instructed facial mimicry to ensure that the facial expressions of the android they used were visible, feasible to imitate, and that electromyography (EMG) was working properly. Interestingly, they reported similar results for spontaneous and instructed facial mimicry. In fact, similar to spontaneous facial mimicry, the instructed facial mimicry of the video-recorded android was less intense than the one directed to the video-recorded human. This result brought us to hypothesize that instructed facial mimicry might be somehow linked to spontaneous facial mimicry. To deepen our understanding of the relationship between instructed and spontaneous facial mimicry, in this paper, \textit{we explore whether spontaneous facial mimicry can be predicted by people's ability to accurately reproduce the dynamics of an agent's facial expressions of the six basic emotions upon instruction to do so}. Moreover, \textit{we study whether artificial agents' embodiment and level of humanlikeness can affect instructed facial mimicry in a way that is analogous to spontaneous facial mimicry}. Should instructed facial mimicry be found to significantly predict spontaneous facial mimicry, it could be used as an \textit{explicit} cue of people's social tuning with an artificial agent and could act as proxy of spontaneous facial mimicry.

\section{Research Questions and Hypotheses}
\label{sec:RQs}

In this work, we explore the influence of {embodiment} and {humanlikeness} on people's spontaneous and instructed mimicry of artificial agents' facial expressions of the six basic emotions. Based on Hofree et al. \cite{hofree2014bridging} and Mattheij et al. \cite{mattheij2013vocal, mattheij2015mirror}, we chose three embodiments for this study: a video-recorded robot, a physical robot, and a virtual agent. Furthermore, we added a fourth condition acting as a control in which participants observed the facial expressions of a video-recorded human. To change the artificial agents' level of humanlikenss, we manipulated their facial features to resemble those of a characterlike face, a humanlike face, and a face that includes features from both of them (i.e., a morph). Humanlikeness was chosen as an independent variable in our study not only because Hofree et al. \cite{hofree2014bridging} found it to be salient for facial mimicry, but also since it is known to influence people's perceptions of an agent's anthropomorphism, social presence, and uncanniness \cite{mori2012uncanny}, \cite{katsyri2015review}, which are perceptual dimensions that in turn affect liking and rapport. Our first group of research questions (RQ1a - RQ1c) concerns spontaneous facial mimicry:

\begin{itemize}
    \item[\textbf{RQ1a}] To what extent does the humanlikeness of artificial agents influence people's spontaneous facial mimicry?
    \item[\textbf{RQ1b}] To what extent does the embodiment of artificial agents influence people's spontaneous facial mimicry?
    \item[\textbf{RQ1c}] Does spontaneous facial mimicry differ between artificial and human agents?
\end{itemize}

Our second group of research questions (RQ2a - RQ2c) revolves around instructed facial mimicry. In previous work \cite{paetzel2017investigating}, we investigated how well people were able to reproduce the dynamics of a laughter performed by an artificial agent that they were explicitly instructed to mimic. In this paper we focus on facial expressions of the six basic emotions instead. Here, we aim to understand whether the agents' embodiment and humanlikeness can affect instructed facial mimicry similar to how they affect spontaneous facial mimicry. Therefore, we pose the following research questions:

\begin{itemize}
\item[\textbf{RQ2a}] To what extent does the humanlikeness of artificial agents influence people's ability to mimic their facial expressions as accurately as possible when instructed to do so?
\item[\textbf{RQ2b}] To what extent does the embodiment of artificial agents influence people's ability to mimic their facial expressions as accurately as possible when instructed to do so?
\item[\textbf{RQ2c}] Does instructed facial mimicry differ between artificial and human agents?
\end{itemize}

The ultimate aim of our research is to inform the development of implicit and explicit behavioral measures that can extend or replace questionnaire-based investigations of the perception of artificial agents.
Previous work has already highlighted that spontaneous facial mimicry signals liking in first acquaintances \cite{chartrand1999chameleon, kulesza2015face} and rapport in established relationships \cite{hess1995intensity, fischer2012emotional}. Liking and rapport are complex constructs known to be influenced by factors such as the appearance and embodiment of an agent \cite{perugia2021ICanSeeIt, paetzel2020persistence, paetzel2021influence}. In this study, besides understanding the role of embodiment and humanlikeness in facial mimicry, we aim to gain more insights on the relationship between spontaneous facial mimicry and a few of the perceptual dimensions known to influence rapport and liking:

\begin{itemize}
\item[\textbf{RQ3}] To what extent can spontaneous facial mimicry predict the agent's perceived social presence, anthropomorphism, uncanniness, and likability?
\end{itemize}

From the related literature, we know that the occurrence of spontaneous facial mimicry can be an important predictor of the rapport people build with a human or artificial interaction partner. However, due to occlusions of the face and the subtlety of the mimicked facial expressions, it is often difficult to capture and quantify spontaneous facial mimicry in natural settings and more complex interactions. In these contexts, instructed facial mimicry could act as a proxy of spontaneous facial mimicry and could be used in place of a questionnaire as an \textit{explicit} indirect cue of liking and rapport. Our fourth research question is thus concerned with the relation between instructed and spontaneous facial mimicry:

\begin{itemize}
\item[\textbf{RQ4}] To what extent does instructed facial mimicry predict spontaneous facial mimicry?
\end{itemize}

Based on related studies performed by Hofree et al. \cite{hofree2014bridging}, Chartrand and Bargh \cite{chartrand1999chameleon}, Kulesza et al. \cite{kulesza2015face}, and Hess et al. \cite{hess1995intensity}, we expected that: 

\begin{itemize}[labelwidth=1cm]
\item[\textbf{(H1)}] Physically embodied, co-present, humanlike artificial agents elicit higher \textit{spontaneous facial mimicry} with respect to virtually embodied, non-co-present, non-humanlike artificial agents.
\item[\textbf{(H2)}] Physically embodied, co-present, humanlike artificial agents elicit higher \textit{instructed facial mimicry} with respect to virtually embodied, non-co-present, non-humanlike artificial agents.
\item[\textbf{(H3)}] Spontaneous facial mimicry positively predicts people's evaluations of the agents' \textit{anthropomorphism}, \textit{social presence}, and \textit{likability}, and negatively predicts their perceived \textit{uncanniness}.
\item[\textbf{(H4)}] Instructed facial mimicry positively predicts spontaneous facial mimicry.
\end{itemize}

\section{Methodology}

\begin{figure}[t!]
\vspace{0.2cm}
\centering
\includegraphics[width=0.9\columnwidth]{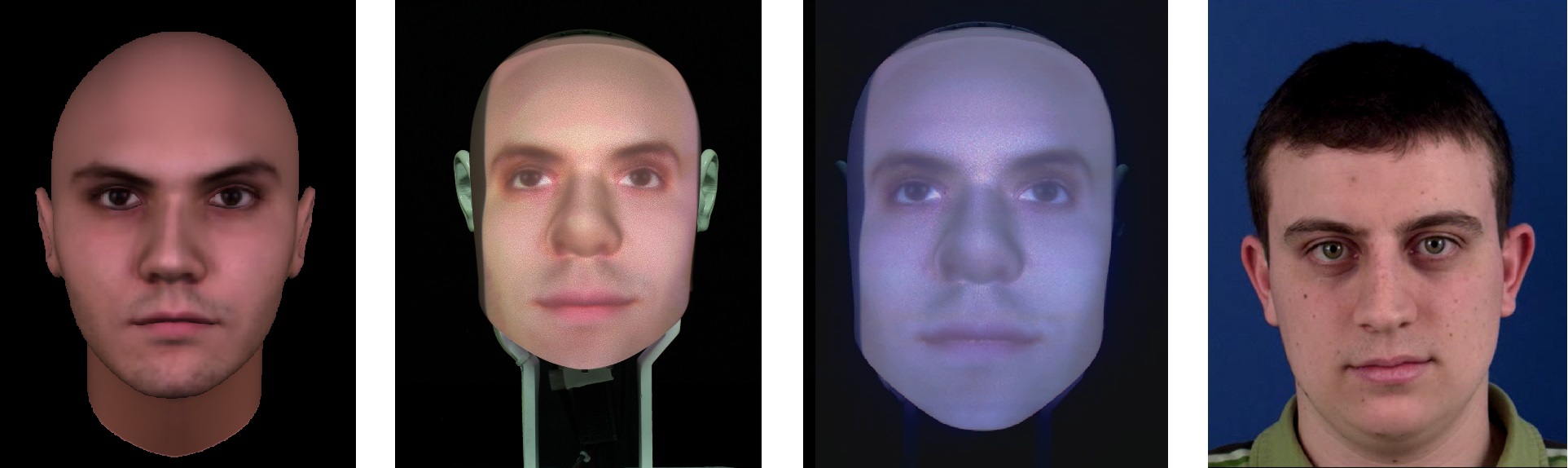}
\caption{Different types of embodiments used in the experiment. From left to right: a virtual agent; a physical Furhat robot; and a video recording of the Furhat robot. The rightmost picture shows the human video used as control condition. }
\label{fig:embodiment}
\end{figure}

\begin{figure}[b!]
\vspace{0.2cm}
\centering
\includegraphics[width=0.75\columnwidth]{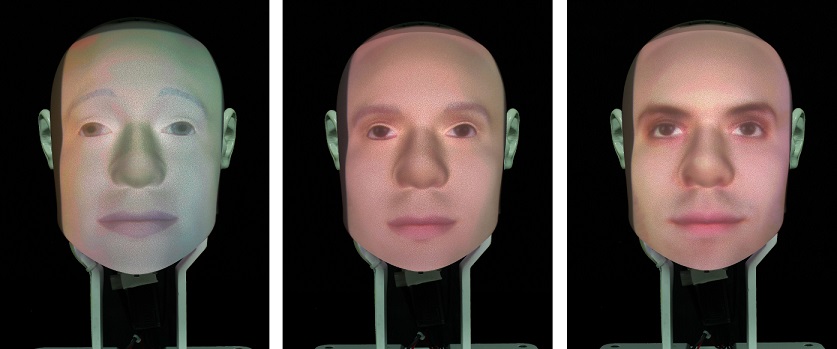}
\caption{Levels of humanlikeness used in the experiment. Left: characterlike; right: humanlike; center: morph.}
\label{fig:textures}
\end{figure}

Our study followed a 3x3 mixed experimental design with:
\begin{itemize}
    \item \textit{Embodiment} as within-subject variable with three types of embodiment: a virtual agent, a physical Furhat robot \cite{almoubayed2012furhat}, and a video-recording of the Furhat robot (\cf~Fig. \ref{fig:embodiment})
    \item \textit{Humanlikeness} as between-subject variable with three levels of humanlikeness: \textit{humanlike}, \textit{characterlike} and a \textit{morph} between the humanlike and the characterlike (\cf~Fig. \ref{fig:textures})
\end{itemize} 
Furthermore, we included a control condition in which participants observed a video-recorded human (\cf fig. \ref{fig:embodiment}). This control condition was the same across all levels of the agent's humanlikeness.

The experimental design was informed by Kulesza et al. \cite{kulesza2015face} and consisted of two parts. In the first part, each participant was asked to observe the facial expressions of the agents and identify which of the six basic emotions they displayed (i.e., happiness, sadness, surprise, anger, fear, disgust). In the second part, which occurred after a 5-minute break, participants were explicitly told that the accuracy of mimicry could improve emotion recognition. Consequently, they were instructed to observe the facial expressions corresponding to the six basic emotions performed by the same agents (in randomized order), mimic them as closely as possible, and identify them only after they finished mimicking. 
The first part of the experiment allowed us to study spontaneous facial mimicry, the second part to investigate instructed facial mimicry. Participants were video-recorded during both parts of the study.

\begin{figure*}[t!]
\vspace{0.2cm}
\centering
\includegraphics[width=0.15\textwidth]{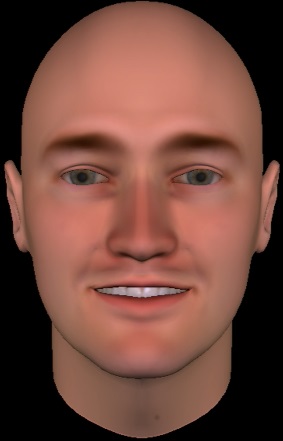}
\hspace{0.05cm}
\includegraphics[width=0.15\textwidth]{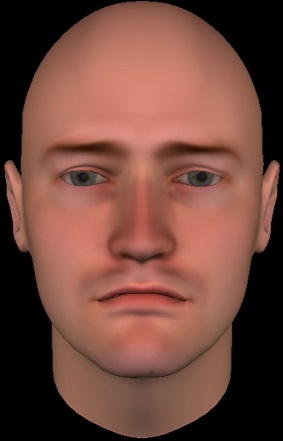}
\hspace{0.05cm}
\includegraphics[width=0.15\textwidth]{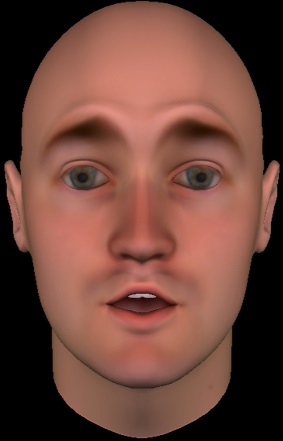}
\hspace{0.05cm}
\includegraphics[width=0.15\textwidth]{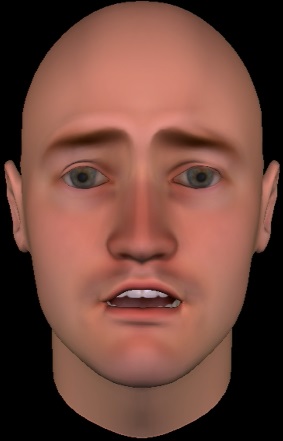}
\hspace{0.05cm}
\includegraphics[width=0.15\textwidth]{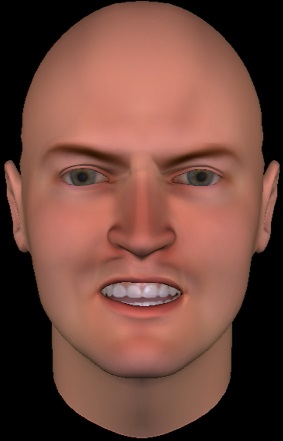}
\hspace{0.05cm}
\includegraphics[width=0.15\textwidth]{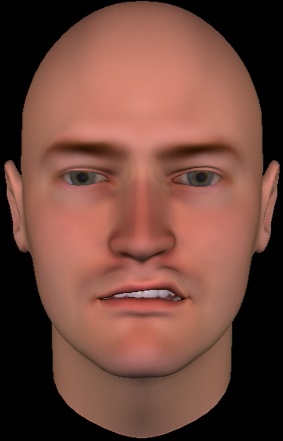}
\caption{Facial expressions of the six basic emotions. From left to right: happiness, sadness, surprise, fear, anger and disgust.}
\label{fig:expressions}
\end{figure*}

Each participant observed a set of facial expressions performed by the video-recorded human and the three artificial agents. All three artificial agents had the same level of humanlikeness but differed in their embodiment. Each set of facial expressions was composed of expressions of the six basic emotions performed twice by each agent. Within each set, the order of presentation of the stimuli was randomized, and no two facial expressions of the same type occurred one after the other. The order of presentation of the artificial and human agents was shuffled using Latin Squares. In total, each participant observed 48 facial expressions for each part of the study. Emotional facial expressions were presented in short sequences of 5 seconds including onset, apex and offset, without vocalizations nor head movements (\cf~Fig. \ref{fig:expressions}). 

\subsection{Participants}
We recruited 46 participants from an international study program in Computer Science at Uppsala University. Participants had at least a high school degree and came from a diverse geographic background (44.4\% Swedish). The 46 participants were randomly allocated to the three conditions corresponding to the different levels of humanlikeness of the artificial agents: \textit{characterlike} (N=15; 11 male; 4 female; 0 other/prefer not to say), \textit{humanlike} (N=16; 13 male; 3 female; 0 other/prefer not to say), and \textit{morph} (N=15; 12 male; 3 female; 0 other/prefer not to say). Due to a misunderstanding of the study task, we excluded the data of one male participant from the humanlike condition. The final sample of participants had a mean age of 26.16 years (\sd=4.37) and was composed of 10 people identifying themselves a female and 35 as male. 
None of the participants had previously interacted with the Furhat robot.

\subsection{Embodiment and Humanlikeness}
\label{sec:embodiment}

As a robot, we chose the Furhat platform \cite{almoubayed2012furhat}. Furhat is a blended robot head consisting of a rigid mask on which a facial texture is projected from within. We chose the Furhat robot for this experiment as its virtual face allowed us to easily alter facial features and design smooth and noiseless facial expressions. 

We designed three different facial textures for the artificial agents. The humanlike face was created from pictures of a real human face using the FaceGen Modeller \footnote{\url{https://facegen.com/modeller.htm}}. The characterlike face was the standard Furhat face with sketched “drawing-like” lips and eyebrows. Finally, the morph face was created by blending the humanlike and characterlike skin textures in the Paint.NET digital photo editing package. The three different textures we applied to the artificial agents were selected from a set of 28 faces tested in a pre-study on Amazon Mechanical Turk (AMT). 
Since initial experiments with the Furhat robot found the face mask without any projection to be perceived as male and dominant \cite{paetzel:icsr}, we limited the set of stimuli to male faces. The same texture we used for the Furhat robot was also utilized to create the virtual agent's face. The video-recorded robot was obtained by recording the physical Furhat. 
For the human condition, instead, we selected the video-recordings of a male person from the MUG database \cite{aifanti2010mug}. 

\subsection{Synthesis of Facial Expressions}

The human in the MUG database was video-recorded while performing the facial expressions of the six basic emotions following the Facial Action Coding System (FACS, \cite{EF02, ekman1978facial}) and an onset-apex-offset temporal scheme.
We designed the facial expressions of the artificial agents by replicating the dynamics of the human video recording as closely as possible. Unfortunately, as in Furhat's IrisTK animation system \cite{skantze2012iristk}, some facial Action Units (AUs) are combined and cannot be controlled separately, the facial expressions of the human and those of the artificial agents slightly differed. An expert trained in the FACS ensured that the final set of stimuli for the artificial agents was still following the FACS' guidelines. Furthermore, an online pre-study conducted on AMT found no systematic difference between the artificial synthesis and the human stimulus in terms of emotion recognition.

\subsection{Experimental Setup}

\begin{figure}
    \centering
    \includegraphics[width=0.5\textwidth]{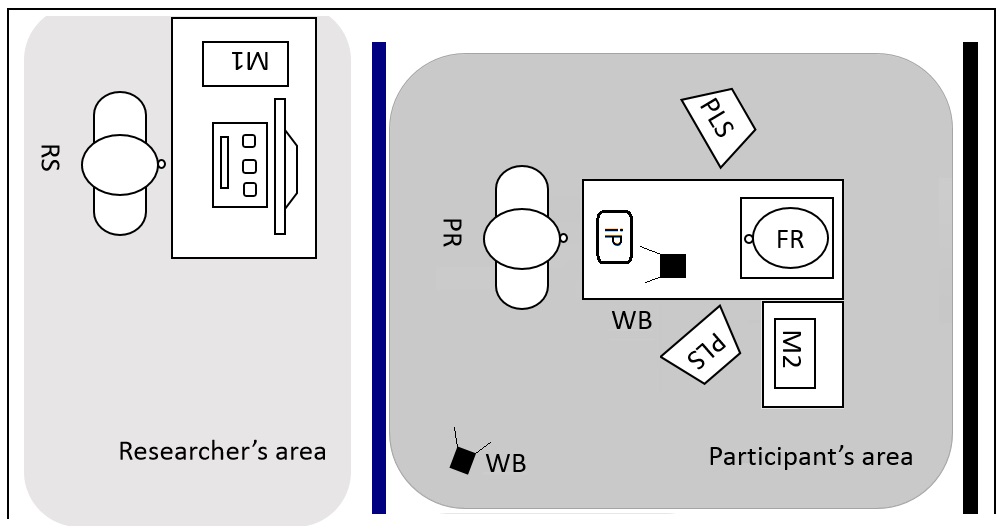}
    \includegraphics[width=0.41\textwidth]{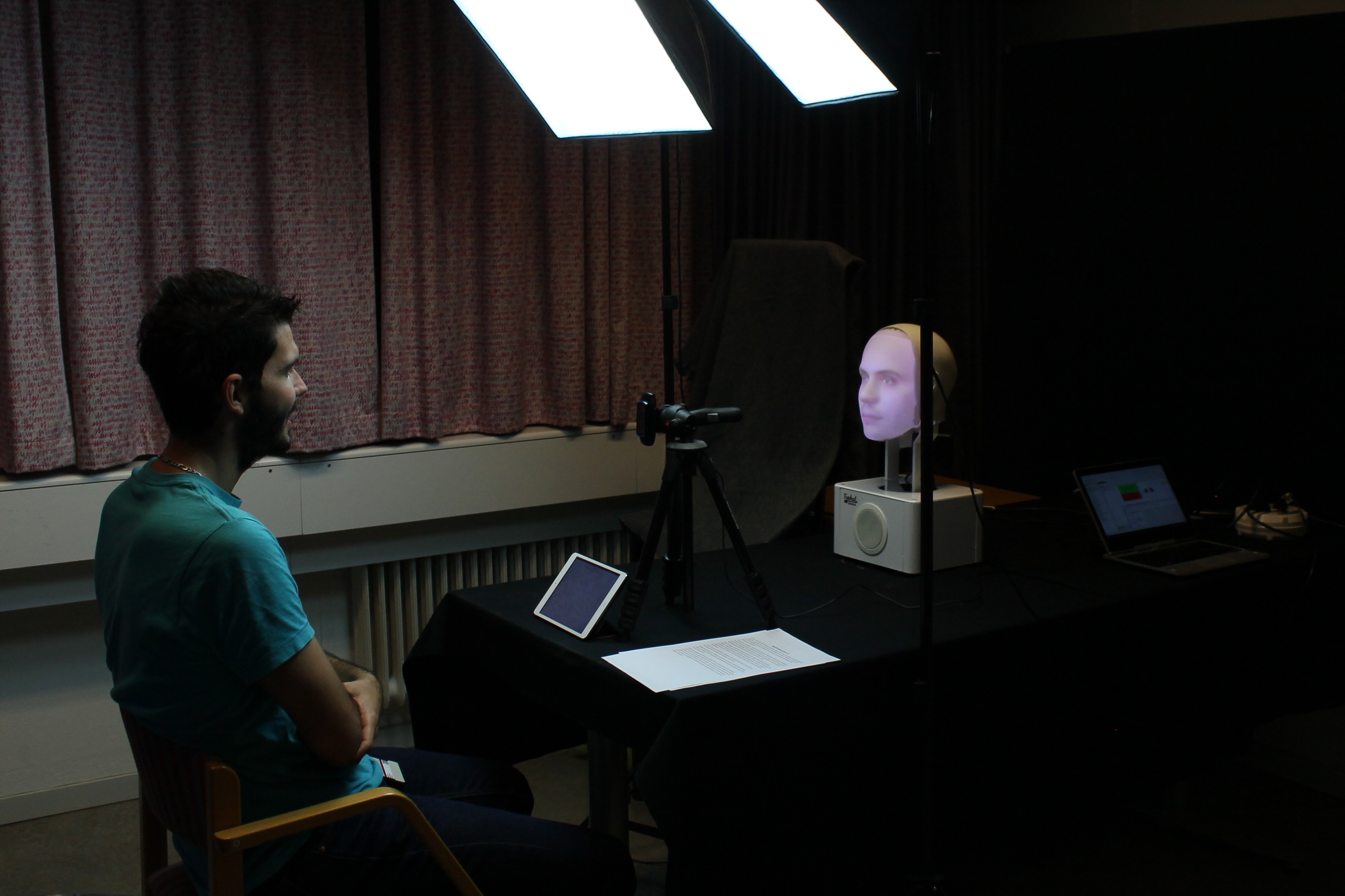}
    \caption{Left: the experimental setup. Right: A participant in the participant area.}
    \label{fig:setup}
\end{figure}

The experimental sessions took place in a private laboratory room at Uppsala University (\cf~Fig. \ref{fig:setup}).
To grant a feeling of privacy and an even background for the video-recordings, the participant's area was separated from the researcher's area by a blue curtain. Black curtains positioned behind the Furhat robot (FR) and the screen displaying the other agents ensured a good visibility of the agents from the participants' perspective. Uniform lighting for the recordings was guaranteed through a professional lighting system (PLS) composed by two lamps. These were the only light sources in the experiment space. As both Furhat and the screen displaying the agents were sources of light themselves, the dark environment ensured a good visibility of the facial expressions.

The participant (PR) was sitting in the participants' area at a distance of about 100 cm from the Furhat robot or the screen. This value falls in the \textit{personal space} of the participants according to Hall \cite{hall_hidden_1969}. The agents were thus close enough to the participants to be properly seen, but not too close to elicit an intimidating feeling. The agents were placed on a table at approximately 100 cm from the ground, which was roughly at eye level for the majority of participants. 
The video-recorded and the virtual agents were presented on a screen in portrait orientation. Their size was calibrated to match the size of Furhat's head.
All embodiments were controlled by a desktop computer (M1). An iPad (iP), placed on the table in front of the participant, was used for answering the questionnaires. 

\subsection{Measures}

\subsubsection{Facial Recordings}
\label{sec:facialrecordings}

To record participants' faces, we used two LOGITECH C920HD PRO webcams (WB) with a 800x600 resolution, operating at 30 fps. The webcams were placed on top of a tripod. One was positioned in front of participants, at approximately 60 cm from them, and slightly on their side to not occlude the stimulus. The second was positioned on the side of the participant (\cf~Fig. \ref{fig:setup}). The webcams were connected to a laptop (M2) which was used to start, stop, and control the video-recordings during the experiment.
Each webcam recorded the entire experimental session with the exclusion of the break between the spontaneous and instructed mimicry trials. Hence, we obtained two video files per camera, participant and session. The video-recordings of the frontal camera  were used to assess participants' mimicry, those of the lateral camera to capture the entire experimental scene. 

\subsubsection{Questionnaires}
Throughout the experiment, four different questionnaires were used. Questionnaire Q1 consisted of a general demographic questionnaire (10 items), the short version of the Big Five personality traits (10 items, \cite{rammstedt2007measuring}) and the Interpersonal Reactivity Index (IRI, 21 questions, excluded personal distress, Cronbach's $\alpha$ between .70 and .78 according to Davis et al. \cite{davis1980multidimensional}). This questionnaire gauged the empathy and personality traits of the participants, and hence was not used to answer this paper's research questions.

Questionnaire Q2 was shown to participants after every facial expression they observed to assess the emotion they recognized in the stimulus. It was composed of the question ``Which of these facial expressions was just displayed?'' with the six basic emotions, ``neutral'' and ``I don't know'' as response options, and the question: ``How certain are you of the selection you made in question 1?'' with a three point Likert scale using the labels: ``Uncertain'', ``Neither nor'', ``Certain''. The response options in the first question were displayed in one of three pre-shuffled orders to prevent a bias towards the first item on the scale.

Questionnaire (Q3) was shown after every embodiment in the first part of the experiment (i.e., spontaneous mimicry trial) to measure participants' perceptions of the agents on four dimensions:
\begin{itemize}
\item \textit{Anthropomorphism} (5 items, 5-point Likert scale), sub-scale from the Godspeed questionnaire by Bartneck et al. \cite{bartneck2009measurement} (Cronbach's $\alpha = .91$ according to Ho and MacDorman \cite{ho2010revisiting}).
\item \textit{Social presence} (8 items, 5-point Likert scale), excerpt from the social presence questionnaire developed by Harms and Biocca \cite{harms2004internal}. Sub-scales: co-presence (2 items, $\alpha = .84$), Attentional Allocation (2 items, $\alpha = .81$), Perceived Affective Understanding (2 items, $\alpha = .86$), Perceived Emotional Interdependence (1 item, $\alpha = .85$) and Perceived Behavioral Interdependence (1 item, $\alpha =. 82$).
\item \textit{Uncanniness and Likability} (10 items, 5-point Likert scale), excerpt from Rosenthal von Der Pütten and Krämer \cite{rosenthal2014design}, sub-scales likability and perceived threat (Cronbach's $\alpha >= 0.82$ for both sub-scales).
\end{itemize}
The order of questions and items remained the same across all embodiments. 

At the end of the experimental session, the experimenter performed a semi-structured interview with the participant. The interview covered potential previous interactions with the Furhat robot, whether participants found aspects in the appearance of one of the characters particularly eerie, and if they had the impression that some of the facial expressions they observed were more difficult to trace back to a specific emotion. This interview was used to gather additional information about the experiment, and was not used to answer any research question present in this paper.

\subsection{Procedure}

After arriving to the lab, participants were informed about the experimental procedure, signed a consent form and answered Q1 on the iPad in front of them. 

For the first part of the experiment, 
participants were asked to first watch the facial expressions displayed by the four agents, which always started and ended with a beep tone, and then indicate which emotions they corresponded to using the questionnaire Q2 displayed on the iPad. Participants were also explained that, once they finished completing Q2 on the iPad and after a pause of about 2 seconds, the agent would automatically display the next facial expression preceded and followed by another beep tone, and the same procedure would be repeated until they had observed all facial expressions.

After participants observed all 12 expressions (2 trials x 6 emotions) for one embodiment, they rated their perception of the observed agent using questionnaire Q3 on the iPad. When neccessary, the experimenter used this lapse of time to switch the physical robot with the screen. Once the participant finished responding to Q3, the stimuli for the subsequent embodiment were shown.
Once participants responded to Q3 for the fourth and final agent of the spontaneous mimicry condition, they were given a five minute break and served refreshments.

For the second part of the experiment, participants were told that research suggests that mimicry increases emotion recognition. Therefore, they 
were asked to perform the same task once again, but this time by first mimicking the facial expression as accurately as possible and then noting down the emotion. The second part of the experiment followed the same procedure of the first part but the embodiments were re-shuffled in order. As Q3 was omitted for the second part of the experiment, participants had a shorter break between embodiments.

At the end of the session, the experimenter conducted the short semi-structured interview. This was followed by a debriefing in which the researcher explained the true nature and objective of the experiment. Participants were informed again that they could request the deletion of their data at any point in time.

\section{Mimicry Processing}
\label{sec:videoanalysis}

The strategy to segment the videos differed between spontaneous and instructed facial mimicry. In the first case, we were interested in understanding whether people mimicked the observed facial expressions or not, whereas in the second case, we were interested in understanding how accurate people were in mimicking the dynamics of the observed facial expressions. This difference in focus is motivated by the different expected magnitudes of spontaneous and instructed facial mimicry. While the former is a subtle response that does not necessarily follow the same dynamics of the expression observed, the latter was expected to be a much stronger and accurate response due to its explicitly imitative nature.

For spontaneous facial mimicry, we annotated the frontal videos of the corresponding trial with the beginning and end of each stimulus in ELAN 5.4. To do so, we used the audible beep tones that marked the start and end of each facial expression of the agents. We then used the minutes obtained from the annotation to automatically cut the original video into shorter snippets using ffmpeg\footnote{https://ffmpeg.org/}. 
To properly divide the instructed mimicry episodes, instead, we first manually identified the initial and final mimicry frames for each stimulus by closely examining the participant's AU activation, and then we cut the original video a second before and after these frames. This process ideally led to 96 individual video snippets per participant, 48 for spontaneous and 48 for instructed facial mimicry. 

Once the data were segmented, we deployed an automatic AU intensity detector to recognize which muscles of the participants' face were activated in each video snippet of the spontaneous and instructed facial mimicry trials (\cf~subsection \ref{sec:detectionAU}). Then, in the case of spontaneous facial mimicry, we checked the AU time series to understand whether or not the target AU or combination of AUs amounting to each facial expression was active for a given lapse of time (\cf~subsection \ref{sec:proc_spontaneous}). In the case of instructed facial mimicry, instead, we used the AU time series to perform a Cross-Recurrence Quantification Analysis (CRQA, \cite{varni2017computational}) as detailed in Section \ref{sec:proc_instructed}.

\subsection{Detection of AU Activation}
\label{sec:detectionAU}

\begin{figure*}
\centering
\includegraphics[width=0.8\linewidth]{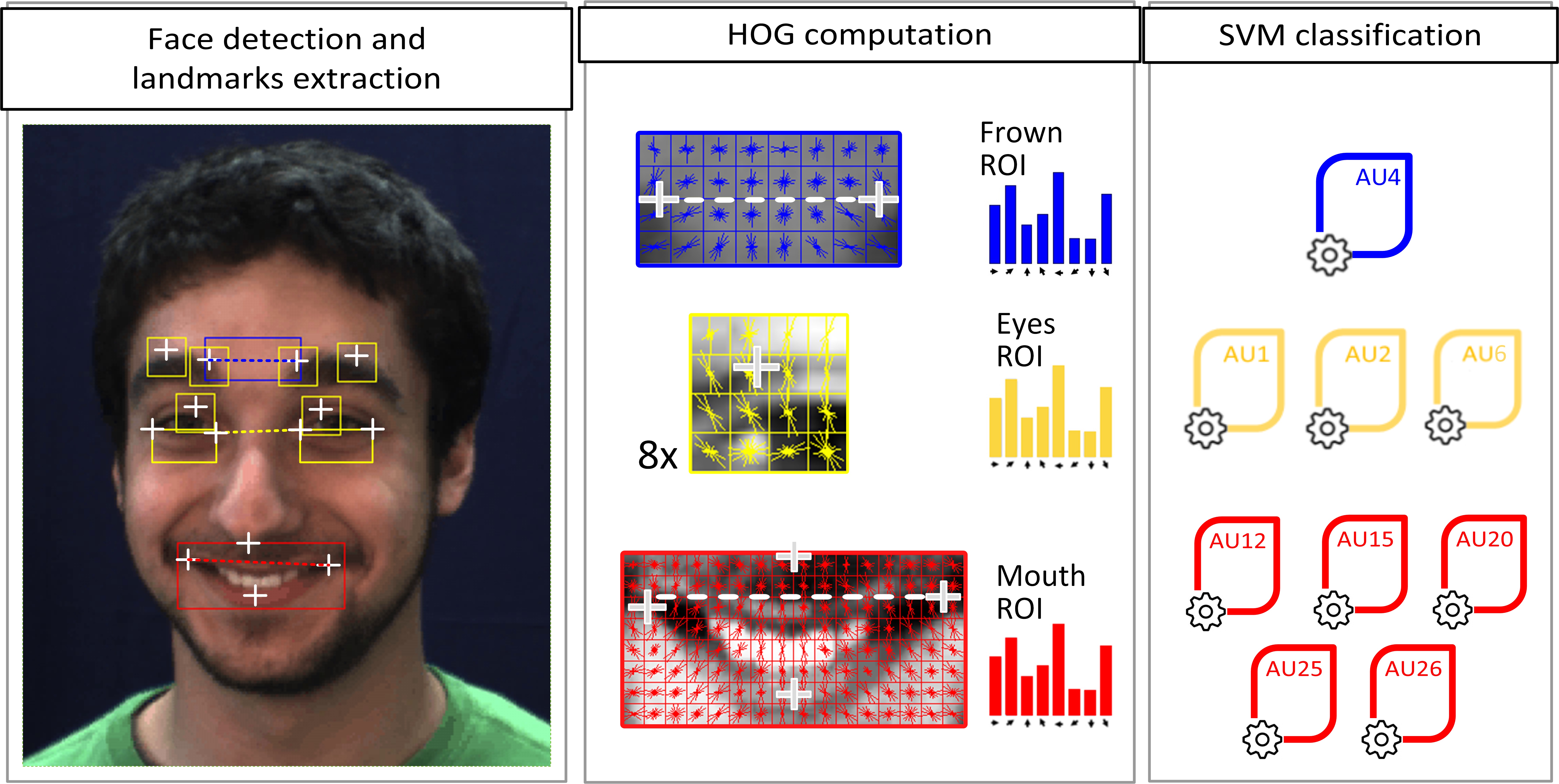}
\caption{AU intensity detection pipeline. The white crosses represent the facial landmarks extracted from the face, while the dashed lines link the landmarks used for aligning each ROI.}
\label{fig:AUdetector_pipeline}
\end{figure*}

The AU intensity detector used in this work is presented in Hupont and Chetouani \cite{hupont2019} and follows the pipeline shown in Fig.~\ref{fig:AUdetector_pipeline}. In a first step, it segments the face of the person from the whole input image and extracts a set of facial landmarks.
Face segmentation is carried out by means of the Viola and Jones' Haar Cascade algorithm \cite{VJ04}. The landmarks (14 white crosses in Figure~\ref{fig:AUdetector_pipeline}) are extracted with the Intraface library introduced by Xiong and Torre \cite{XD13}.
On the basis of the facial landmark positions, three rectangular facial Regions of Interest (ROIs) are then defined and features of Histogram Oriented Gradients (HOG, \cite{HOG}) are computed for each one of them. The ROIs used in our pipeline are:

\begin{itemize}
\item \textbf{Frown ROI} (used for AU4 model): This ROI is located around the inner eyebrow landmarks, which are also used for alignment purposes.
\item \textbf{Eyes ROI} (AU1, AU2 and AU6): This ROI is made up of 8 patches located around the inner eyebrows, the middle eyebrows and the eye landmarks. ROI alignment is performed using inner eye corners. The final descriptor results from the concatenation of the 8 HOG descriptors.
\item \textbf{Mouth ROI} (AU12, AU15, AU20, AU25 and AU26): This ROI is bounded by the nose center, the two lip corners and the lower lip. Alignment is done with respect to the lip corner positions.
\end{itemize}

Finally, the classification of each AU in terms of intensity is performed by an individually pre-trained Support Vector Machine (SVM) model using its corresponding ROI features as input. The SVM models were trained on the large-scale DISFA facial action database \cite{DISFA}. Each model detects the activation of its corresponding AU in terms of six intensity categories, which are, according to Ekman's taxonomy \cite{EF02}: ``N" (neutral), ``A" (trace), ``B" (slight), ``C" (marked), ``D" (severe) and ``E" (maximum). The AU detector achieved an overall Intraclass Correlation Coefficient \textit{ICC(3,1)} of 0.73, which is within state-of-the-art performances in the task of AU intensity detection.

The AU intensity time series was low-pass filtered through a centered moving average filter with a window size of 10 samples (33.3ms). This filtering was applied to both the spontaneous and instructed facial mimicry time series. 
Moreover, the duration of time for which each AU was activated was also computed. For instructed facial mimicry, the first and the last 30 samples corresponding to the 1 second buffer left before and after the initial and final mimicry frames were removed in the final time series.

\subsection{Processing of Spontaneous Facial Mimicry}
\label{sec:proc_spontaneous}

To assess spontaneous facial mimicry, we divided the AU time series into two time intervals. The first time interval spanned from 0 to 1000 ms after stimulus onset and encompassed quick mimicry responses occurring at a subperceptual level, which Dimberg et al. \cite{dimberg2000unconscious} call Rapid Facial Reactions (RFR). The second time interval ranged from 1000 to 5000 ms after stimulus onset and comprised facial mimicry responses occurring at a more conscious level, which we call Controlled Facial Reactions (CFR).

To consider a facial expression as mimicked at each time interval (RFR, CFR), we checked whether the AU or combination of AUs corresponding to the target facial expression (\cf~Table \ref{table:Emotions} based on Ekman et al. \cite{ekman1978facial}) was active for at least 3 consecutive frames (100 ms). The activation was coded as 0 (not activated) or 1 (activated) and the intensity of the activation was not considered for this analysis as we expected the intensity of spontaneous facial mimicry to be low. We chose the threshold of 100 ms based on Ito et al. \cite{ito2004fast}, who defined this as the shortest period of time a muscle can take to move.
To perform the statistical analyses, we calculated the percentage of spontaneous facial mimicry for RFR and CFR. This value was obtained \textit{per embodiment} by dividing the number of trials in which the participant mimicked the facial expressions by the number of valid video snippets for that embodiment.
Since in the spontaneous mimicry part of the study, participants were not explicitly asked to mimic the facial expressions they observed, in some snippets their faces were occluded, out of frame, or not recognizable by the AU intensity detector. These snippets were excluded from the final analyses. If more than half of the snippets of a particular embodiment were missing, we also excluded the other valid snippets from the a final analysis. Overall, this led to the exclusion of a total of 465 snippets for RFR (22\%) and 394 for CFR (18\%), and left us with 1695 valid snippets for RFR, and 1766 for CFR.

\begin{table}[]
\centering
\caption{AUs or combination of AUs used to detect the spontaneous mimicry of the facial expressions of the six basic emotions (based on \cite{ekman1978facial})}
\vspace{0.2cm}
\begin{tabular}{ll}
\textbf{Emotion} & \textbf{Action Units (AUs)}  \\
Anger            & AU4\\
Disgust          & AU4 + AU25 \\
Fear             & AU20, AU1 + AU2 + AU4 \\
Happiness        & AU6, AU12, AU6 + AU12 \\
Sadness          & AU1, AU15, AU1 + AU4 \\
Surprise         & AU26, AU1 + AU2
\end{tabular}
\label{table:Emotions}
\end{table}

\subsection{Analysis of Instructed Facial Mimicry}
\label{sec:proc_instructed}

\begin{figure}[b!]
\vspace{0.2cm}
\centering
\includegraphics[width=0.4\columnwidth]{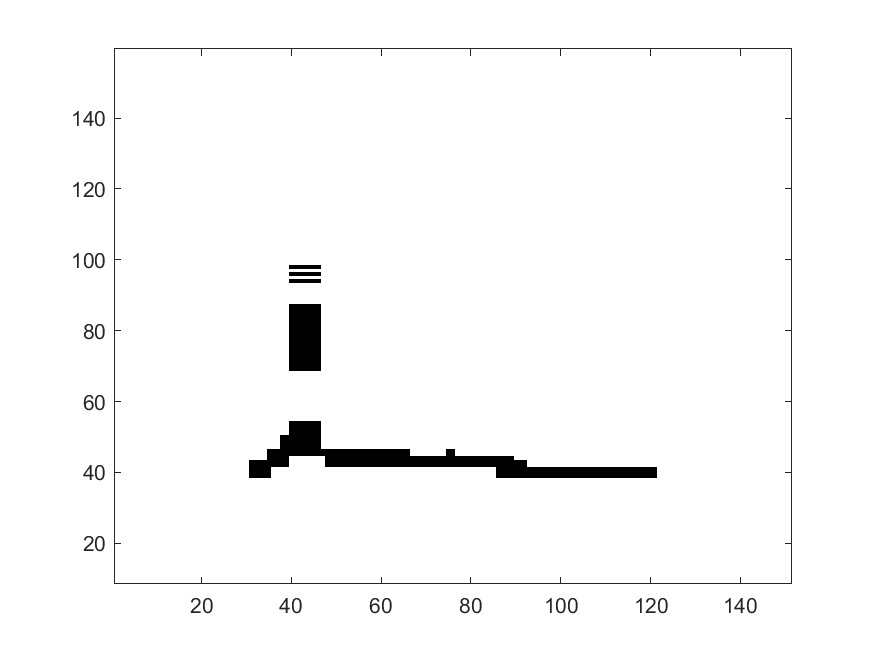}
\includegraphics[width=0.4\columnwidth]{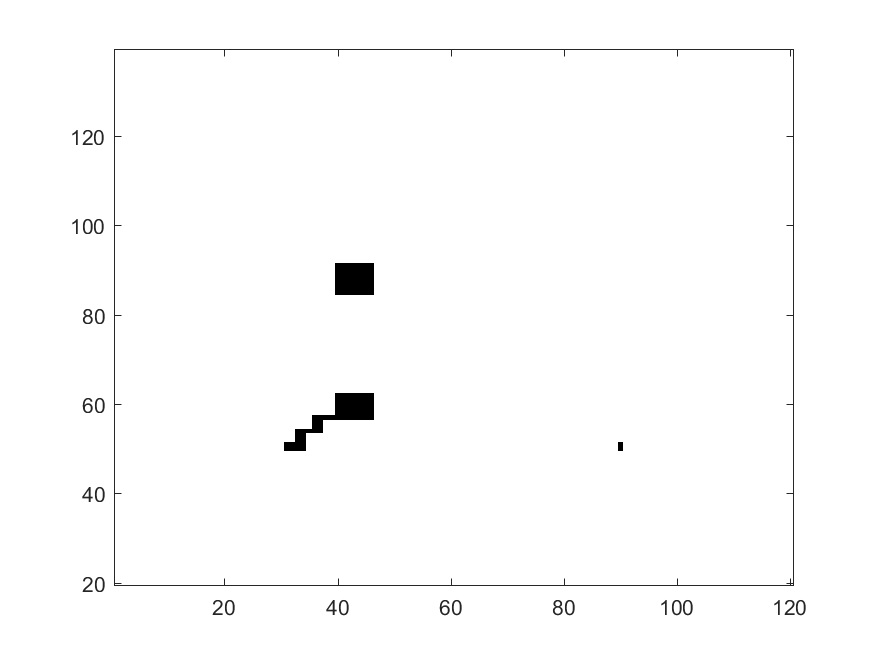}
\caption{Cross-Recurrence Plot of two different participants intentionally mimicking the facial expressions of anger displayed by the virtual agent (left) and the physical Furhat (right).}
\label{fig:crp}
\end{figure}

In order to accurately assess the dynamics of facial expressions, we performed a CRQA analysis
\cite{Marwan07}. CRQA is a technique enabling a quantitative measure of the graphical patterns occurring in a Cross-Recurrence Plot (CRP, \cf~Fig. \ref{fig:crp}). CRP is a plot looking at the times at which the features of a dynamical system \emph{recur} (i.e., it is \emph{close}) to features of another dynamical system. In this study, the two dynamical systems were the user and the artificial agents / video-recorded human, and the features were the AU intensities. 

A CRP can be displayed as a square / rectangular black and white area spanned by two time series describing two systems. Black points correspond to the times at which the two systems co-visit the same area in the feature space, whereas white points correspond to the times at which each system runs in a different area. A CRP is expressed by the following \emph{cross-recurrence matrix} (CR) :
\begin{equation}
	CR_{i,j}^{\vec{f1},\vec{f2}}(\epsilon)=\Theta(\epsilon-\|\vec{f1_i}-\vec{f2_j}\|), \; i=1...N, j=1...M
\end{equation}
where $\vec{f1}$ and $\vec{f2}$ $\in$ ${\rm I\!R}^d$ are the d-dimensional time series of the two systems having N and M samples, respectively; $\epsilon$ is the threshold to claim closeness between two points, $\Theta (.)$ is the Heaviside function and $\|.\|$ is a norm. 
In this study, $\vec{f1}$ and $\vec{f2}$ $\in$ ${\rm I\!R}^3$ are the time series of the AU intensities of the human and the artificial agents / video-recorded human over N samples. The threshold $\epsilon$ was set to 2 expressing that there was a match only when the `distance' between the intensities of corresponding AUs was less than two. The norm used was the Manhattan distance. 

CRPs can be analyzed through the Cross-Recurrence Quantification Analysis (CRQA) that enables to extract quantitative information from the black and white patterns appearing in the plot (see \cite{Marwan07} for a complete survey). Typical patterns are: single isolated points, periodical diagonal lines, and vertical / horizontal lines. These patterns are hints of randomness, periodicity and laminar states of the dynamics of the system. In this study, we focused on the following CRQA measures (\cite{Marwan07}): 

\vspace{0.2cm}\noindent\emph{Cross-Recurrence Rate ($cRR$)}\\
The Cross-Recurrence Rate is defined as:
\begin{equation}
cRR (\epsilon)= \frac{1}{N^2}\sum\limits_{i,j=1}^{N}CR_{i,j}(\epsilon)
\end{equation}

\noindent and measures the density of recurrence points in a CRP. It corresponds to the ratio between the number of the matrix elements shared by the participant and the artificial agents / video-recorded human and the number of available elements (i.e. all the elements in the matrix).  Here, $cRR$ represents the overall extent to which the human and the artificial agent / recorded human were activating the same AUs at a similar level. 
This measure alone, however, even if it is a first measure to address mimicry, does not provide any information about how mimicry unfolds over time. To extract information about that, several other CRQA measures were computed:

\vspace{0.2cm}\noindent\emph{Average diagonal lines length (L) and maximum diagonal line length ($L_{max}$)} \\
$L$ represents the average length of a recurrent trajectory in a CRP. It is defined as:
\begin{equation}
    L=\frac{\sum_{l=l_m}^{N}lP(l)}{\sum_{l=l_m}^{N}P(l)}
\end{equation}
where $l_{m}$ is the minimal diagonal length to be taken into account, and $P(l)$ is the histogram of the diagonal lines. The minimal diagonal length was set to 8 samples, i.e. around 250 ms \cite{FASEL2003259}.  
The value of L expresses how stable a recurrent trajectory is. Here high values of $L$ correspond to long, almost identical portions of AU intensities of the human and the artificial agents over time.  Moreover, the length $L_{max}$ of the longest diagonal line in the CRP was computed. A large value of $L_{max}$ shows a slow divergence of the AUs’ intensity trajectories. 

\vspace{0.2cm}\noindent\emph{Determinism (DET)} \\
As a fourth and last measure, the determinism was computed. It is defined as: 
\begin{equation}
    DET = \frac{\sum_{l=l_m}^{N}lP(l)}{\sum_{l=1}^{N}P(l)}
\end{equation}

\noindent It measures the percentage of the cross-recurrence points forming diagonal lines (of at least length $l_{m}$) computed with respect to all the cross-recurrence points in the CRP. $DET$ ranges in $[0,1]$ and it is a hint of the predictability of the system (when $DET=0$ the systems is stochastics, when $DET=1$ it is periodic). In this study, high values of $DET$ were expected to be found during good mimicry episodes. 

While participants paid more attention to stay in frame during the instructed mimicry phase, we still had to exclude snippets due to occlusions and errors of the AU intensity detector. If there were only 5 or less valid snippets for a particular embodiment and participant, these were removed from the final analysis. Overall, we excluded a total of 209 snippets (9\%) and were left with 1951 valid snippets for the analysis of instructed mimicry. For the statistical analysis, we calculated the average $cRR$, $L_{max}$, $L_{avg}$ and $DET$ of each participant across all valid trials associated with one embodiment.

\section{Results}

In the remainder of the paper, we use: (1) \textit{social presence} to refer to the dependent variables co-presence, attentional allocation, perceived affective understanding, perceived emotional interdependence, and perceived behavioral interdependence; (2) \textit{perception of the agent} to refer to the dependent variables anthropomorphism, likability, and perceived threat; (3) \textit{emotion recognition} to refer to the dependent variables percentage of correctly recognized emotions and average confidence in the recognized emotion; (4) \textit{spontaneous facial mimicry} to refer to the percentage of spontaneous facial mimicry for rapid facial reactions (RFR) and controlled facial reactions (CFR); (5) \textit{instructed facial mimicry} to refer to the average (avg) $cRR$, avg $L$, avg $L_{max}$, and avg $DET$.

For the two manipulation checks (MC1 and MC2) and the preliminary analyses (PA), and for answering RQ1 and RQ2, we performed separate 3x3 repeated measures ANOVAs with humanlikeness as \textit{between-subject factor} (i.e., humanlike, characterlike, morph), embodiment as \textit{within-subject factor} (i.e., virtual agent, physical robot, and video-recording of the physical robot) and (i) social presence (MC1), (ii) perception of the agent (MC2), (iii) emotion recognition (PA), (iv) spontaneous facial mimicry (RQ1) and (v) instructed facial mimicry (RQ2) as \textit{dependent variables}. All p-values that we report in the post-hoc analyses are Bonferroni corrected to account for multiple tests.

For MC2, PA, RQ1, and RQ2, we also ran follow-up 2x3 repeated measures ANOVAs with humanlikeness as a between-subject factor (humanlike, characterlike, morph), \textit{artificiality of the agent} as a within-subject factor (i.e., artificial agents and human agent), and the same dependent variables. To perform these analyses, we calculated the average value across all three artificial agents on each dependent variable. 
Social presence (MC1) was excluded from this set of analyses since the video-recorded human did not vary in embodiment like the artificial agents. We kept humanlikeness as a between-subject factor to control for eventual effects of the different levels of humanlikeness of the artificial agents on the dependent variables. However, as this effect is already covered by the 3x3 repeated measures ANOVAs, for the sake of brevity, we do not report these results. All the p-values that we report in the post-hoc analyses are Bonferroni corrected to account for multiple tests.

Finally, for answering RQ3 and RQ4, we performed separate regression analyses using spontaneous facial mimicry as a predictor of social presence and perceptions of the artificial agents (RQ3) and instructed facial mimicry as a predictor of spontaneous facial mimicry (RQ4). As RQ3 specifically focused on artificial agents' facial mimicry, we used only the data from the artificial agents to perform the regression analyses. On the contrary, as RQ4 focused on facial mimicry in general and not specifically on artificial agents' mimicry, we included also the data from the human video in the regression analyses.

\subsection{Manipulation Check and Preliminary Analyses}

\subsubsection{Manipulation Check: Social Presence of the Artificial Agents}

The results indicated a significant main effect of embodiment on co-presence, affective understanding, and emotional interdependence (\cf~Table \ref{table:3x3ANOVA} for the complete results). Furthermore, they showed a significant interaction effect of humanlikeness and embodiment on co-presence.

Post-hoc analyses uncovered that the virtual agent was perceived as significantly more co-present than the video-recorded robot ($p=.005$, \cf~Table \ref{table:descriptive_embodiment} for the descriptive statistics), and the physical robot was perceived as significantly more co-present ($p=.005$) than the video-recorded one. No such difference was observed between the virtual agent and the physical robot ($p=1.00$). Moreover, they disclosed that participants perceived their affective understanding of the physical robot to be significantly higher than that of the virtual agent ($p=.045$), while the virtual agent and the video-recorded robot did not differ in terms of perceived affective understanding ($p=.255$), and neither did the physical robot and the video-recorded one ($p=1.00$). Finally, participants perceived significantly higher emotional interdependence with the physical robot with respect to both the virtual agent ($p=.021$, \cf~Table \ref{table:descriptive_embodiment} for the descriptive statistics) and the video-recorded robot ($p=.019$). No such difference was present between the virtual agent and the video-recorded robot ($p=1.00$).

Further follow-up post-hoc analyses on the interaction effect of humanlikeness and embodiment on co-presence uncovered that, in the characterlike condition, the virtual agent ($M=4.10$, $SD=.632$) and the physical robot ($M=4.27$, $SD=.729$) were perceived as significantly more co-present than the video-recorded robot ($M=3.70$, $SD=.621$, virtual agent: $p=.026$; physical robot: $p=.028$), but they did not significantly differ in co-presence from each other ($p=1.00$). Likewise, in the morph condition, the virtual agent ($M=3.71$, $SD=1.051$) was perceived as significantly more co-present than the video-recorded robot ($M=3.07$, $SD=.938$, $p=.016$), the physical robot ($M=3.53$, $SD=.930$) was perceived as significantly more co-present than the video-recorded one ($p=.051$), and the virtual agent and the physical robot did not differ from each other ($p=.409$). Interestingly though, in the humanlike condition (virtual agent: $M=3.79$, $SD=.777$; physical robot: $M=4.04$, $SD=.746$; video-recorded robot: $M=4.00$, $SD=.734$), these differences between artificial agents were not present (virtual agent - physical robot: $p=.331$; virtual agent - video-recorded robot: $p=.083$; physical robot - video-recorded robot: $p=1.00$).

\begin{table*}[]
\centering
\caption{\textbf{Results of 3x3 Repeated Measures ANOVAs for Manipulation checks and RQ1 and RQ2}. The significant results are displayed in bold, while the trend effects are presented in italics.}
\vspace{0.2cm}
\begin{tabular}{ | r || r | r | r || r | r | r || r | r | r |}
\hline
& \multicolumn{3}{c ||}{\textbf{Embodiment}} & \multicolumn{3}{c |}{\textbf{Humanlikeness}} & \multicolumn{3}{c |}{\textbf{Embod. x Human.}}\\
\hline
\hline
\textbf{Social Presence} & $F(2,80)$ & $p$ & $\eta p^2$ & $F(2,40)$ & $p$ & $\eta p^2$ & $F(4,80)$ & $p$ & $\eta p^2$\\
\hline
Co-presence &\textbf{ 7.878} & \textbf{.001} &\textbf{.165} & \textit{2.719} & \textit{.078} & \textit{.120} & \textbf{4.036} &\textbf{ .005} & \textbf{.168}\\ 
\hline
Att. Allocation & 2.040 & .137 & .049 & .036 & .965 & .002 & 1.377 & .249 & .064 \\ 
\hline
Aff. Understand. & \textbf{3.643} & \textbf{.031} & \textbf{.083} & 1.115 & .338 & .053 & \textit{2.373} &\textit{ .059} & \textit{.106} \\ 
\hline
Em. Interdep. & \textbf{5.864} & \textbf{.004} & \textbf{.128} & 2.157 & .129 & .097 &  .668 & .616 & .032\\ 
\hline
Beha. Interdep. & .630 & .535 & .015 & .750 & .479 & .036 & .725 & .578 & .035\\ 
\hline
\hline
\textbf{Agent's Percept.}& $F(2,80)$ & $p$ & $\eta p^2$ & $F(2,40)$ & $p$ & $\eta p^2$ & $F(4,80)$ & $p$ & $\eta p^2$\\
\hline
Anthropomorph. & \textbf{15.587} & $<\textbf{.001}$ & \textbf{.280 }& \textbf{3.399} & \textbf{.043} & \textbf{.145} & \textit{2.246} & \textit{.071} & \textit{.101}\\ 
\hline
Perceived Threat & \textbf{6.470} & \textbf{.002} & \textbf{.139} & .244 & .785 & .012 & \textit{2.447} & \textit{.053} & \textit{.109}\\ 
\hline
Likability & \textbf{8.361} & \textbf{.001} & \textbf{.173} & 1.776 & .182 & .082 & 1.454 & .224 & .068\\ 
\hline
\hline
\textbf{Emo. Recogn.}& $F(2,80)$ & $p$ & $\eta p^2$ & $F(2,40)$ & $p$ & $\eta p^2$ & $F(4,80)$ & $p$ & $\eta p^2$\\
\hline
Recogn. (Spont.)  & .296 & .745 & .007 & \textbf{4.004 }& \textbf{.026} &\textbf{ .167} & .253 & .907 & .013 \\ 
\hline
Recogn. (Instr.)  & 1.147 & .323 & .028 & \textbf{5.540} & \textbf{.008} & \textbf{.217} & .482 & .749 & .024 \\
\hline
\hline
\textbf{Spont. Mimicry }& $F(2,64)$ & $p$ & $\eta p^2$ & $F(2,32)$ & $p$ & $\eta p^2$ & $F(4,64)$ & $p$ & $\eta p^2$\\
\hline
Freq. RFR & \textbf{9.336} & $<\textbf{.001}$ & \textbf{.226} & 1.002 & .378 & .059 & 1.071 & .378 & .063\\ 
\hline
Freq. CFR & \textbf{4.645} & \textbf{.013} &\textbf{ .127} & .636 & .536 & .038 & 1.566 & .194 & .089\\
\hline
\hline
\textbf{Instr. Mimicry} & $F(2,76)$ & $p$ & $\eta p^2$ & $F(2,38)$ & $p$ & $\eta p^2$ & $F(4,76)$ & $p$ & $\eta p^2$\\
\hline
Avg $cRR$ &  .097 & .908 & .003 & 2.189 & .126 & .103 & .785 & .538 & .040\\ 
\hline
Avg $L$ & .364 & .696 & .009 & .208 & .813 & .011 & .411 & .800 & .021\\ 
\hline
Avg $L_{max}$ & .477 & .662 & .012 & .293 & .748 & .015 & .298 & .878 & .015\\ 
\hline
Avg $DET$  & .219 & .804 & .006 & .187 & .830 & .010 & .784 & .539 & .040\\ 
\hline
\end{tabular}
\label{table:3x3ANOVA}
\end{table*}

\begin{table*}[]
\centering
\caption{\textbf{Descriptive Statistics of the 3x3 Repeated Measures ANOVAs per Embodiment}: Mean (\me) and standard deviation (\sd) of all dependent variables}
\vspace{0.2cm}
\begin{tabular}{ | r || r | r || r | r || r | r |}
\hline
& \multicolumn{2}{c ||}{\textbf{Virtual Agent}} & \multicolumn{2}{c ||}{\textbf{Physical Robot}} & \multicolumn{2}{c |}{\textbf{Video Robot}} \\
& \me & \sd & \me & \sd & \me & \sd\\
\hline
\hline
Co-presence & 3.87 & .832 & 3.95 & .844 & 3.59 & .847\\ 
\hline
Att. Allocation & 4.05 & .837 & 4.20 & .757 & 4.02 & .809 \\ 
\hline
Aff. Understanding & 3.17 & .778 & 3.45 & .625 & 3.27 & .658 \\ 
\hline
Em. Interdependence & 1.79 &  .888 & 2.12 & 1.051 & 1.70 & .832\\ 
\hline
Beha. Interdependence & 2.28 & 1.076 & 2.35 & 1.066 & 2.21 & .914\\ 
\hline
\hline
Anthropomorphism & 2.53 & .834 & 3.10 & .742 & 2.67 & .777 \\ 
\hline
Perceived Threat & 2.10 & .769 & 1.84 & .650 & 1.72 & .524 \\ 
\hline
Likability & 2.22 & .759 & 2.66 & .708 & 2.43 & .787 \\ 
\hline
\hline
Recognized (Spont.)  & .77 & .139 & .79 & .139 & .77 & .129\\ 
\hline
Recognized (Instr.)  & .80 & .151 & .77 & .142 & .79 & .117\\  
\hline
\hline
Freq. RFR & .56 & .165 & .53 & .189 & .66 & .151\\ 
\hline
Freq. CFR & .74 & .174 & .66 & .188 & .73 & .156 \\ 
\hline
\hline
Avg $cRR$ & 13.12 & 7.430 & 13.54 & 6.692 & 13.26 & 6.585 \\ 
\hline
Avg $L$ & 5.52 & 2.987 & 5.29 & 2.754 & 5.66 & 1.656 \\ 
\hline
Avg $L_{max}$ & 6.59 & 3.596 & 6.36 & 3.364 & 6.88 & 3.321 \\ 
\hline
Avg $DET$  & 2.78 & 1.856 & 2.63 & 1.612 & 2.78 & 1.262\\ 
\hline
\end{tabular}
\label{table:descriptive_embodiment}
\end{table*}

\subsubsection{Manipulation Check: Perception of the Agents}

When checking for differences in the perception of the artificial agents across levels of humanlikeness and embodiments, we found a significant main effect of embodiment on anthropomorphism, perceived threat, and likability (\cf~Table \ref{table:3x3ANOVA} for the complete results) and a significant main effect of humanlikeness on anthropomorphism.

Bonferroni-corrected post-hoc analyses revealed that the virtual agent and the video-recorded robot were perceived as less anthropomorphic than the physical robot (both $p<.001$, \cf~Table \ref{table:descriptive_embodiment} for the descriptive statistics), but the virtual agent and the video-recorded robot did not differ in terms of anthropomorphism between each other ($p<.682$). Similarly, in terms of likability, the virtual agent was perceived as less likable than the physical robot ($p=.001$, \cf~Table \ref{table:descriptive_embodiment} for the descriptive statistics) and the video-recorded robot was perceived as less likable than the physical one ($p=.042$). However, the virtual agent and the video-recorded robot did not differ from each other ($p=.291$). Finally, concerning perceived threat, the virtual agent was perceived as more threatening than the video-recorded robot ($p=.001$, \cf~Table \ref{table:descriptive_embodiment} for the descriptive statistics), but no such difference was present between the virtual agent and the physical robot ($p=.104$) and between the video-recorded and the physical robot ($p=.822$) 

With regards to the main effect of humanlikeness, the post-hoc analyses disclosed that humalike artificial agents were perceived as significantly more anthropomorphic than morph artificial agents ($p=.046$, \cf~Table \ref{table:descriptive_humanlikeness} for the descriptive statistics). However, humanlike and characterlike artificial agents ($p=.232$) and characterlike and morph agents ($p=1.00$) did not differ significantly from each other. 

When running the 2x3 ANOVA focusing on the agents' artificiality, we found out that the video-recorded human was perceived as significantly more anthropomorphic ($p<.001$), more likable ($p<.001$), and less threatening ($p<.001$) than the artificial agents (\cf~Table \ref{table:ANOVA_descriptive_artificiality} for the results and the descriptive statistics).

\vspace{0.2cm}\noindent\textbf{Discussion of Manipulation Check.}
As specified in section \ref{sec:mimicryartificial}, the artificial agents and the video-recorded human differed as follows: (i) the physical robot was \textit{artificial, physically embodied}, and \textit{co-present}; (ii) the virtual agent was \textit{artificial, virtually embodied}, and \textit{co-present}; (iii) the video-recorded robot was \textit{artificial, physically embodied}, but \textit{not co-present}; and (iv) the video-recorded human was \textit{natural, physically embodied}, but \textit{not co-present}.
The manipulation checks that we performed were aligned with these differences. Indeed, the video-recorded robot was perceived as significantly less co-present than the virtual agent and physical robot. Furthermore, the artificial agent that was physically embodied and co-present (i.e., the physical robot) was perceived as easier to understand affectively, more anthropomorphic, more likable, and elicited more emotional understanding than the other artificial agents. Finally, the human agent was perceived as more anthropomorphic, more likable, and less threatening than the artificial agents.
As a result, we can state that the manipulation of embodiment worked as expected in this study.

With regards to the manipulation of humanlikeness, the core dependent variable that we expected to change was anthropomorphism. The characterlike and morph robot did not differ in anthropomorphism and neither did the characterlike and humanlike robot. However, in line with our expectations, the humanlike robot was perceived as more anthropomorphic than the morph robot. As a result, we considered the manipulation of humanlikeness only partially successful.
With regards to the manipulation of humanlikeness, it was also very interesting to discover that, when the appearance of the artificial agents was humanlike, the differences in co-presence between the different embodiments ceased to exist. This result seems to suggest that the humanlike appearance has in itself a quality of co-presence that goes beyond the physical instantiation of an artificial agent. 

\begin{table*}[]
\centering
\caption{\textbf{Descriptive Statistics of the 3x3 Repeated Measures ANOVAs per level of humanlikeness}: Mean (\me) and standard deviation (\sd) of all dependent variables}
\vspace{0.2cm}
\begin{tabular}{ | r || r | r || r | r || r | r |}
\hline
& \multicolumn{2}{c ||}{\textbf{Character.}} & \multicolumn{2}{c ||}{\textbf{Humanlike}} & \multicolumn{2}{c |}{\textbf{Morph}} \\
& \me & \sd & \me & \sd & \me & \sd\\
\hline
\hline
Co-presence & 4.02 & .720 & 3.94 & .722 & 3.44 & .722 \\ 
\hline
Att. Allocation & 4.10 & .734 & 4.12 & .737 & 4.05 & .737\\ 
\hline
Aff. Understanding & 3.19 & .550 & 3.48 & .550 & 3.24 & .550\\ 
\hline
Em. Interdependence & 2.20 & .775 & 1.64 & .775 & 1.74 & .775\\ 
\hline
Beha. Interdependence & 2.16 & .910 & 2.52 & .909 & 2.17 & .909\\ 
\hline
\hline
Recognized (Spont.) & .83 & .089 & .74 & .090 & .76 & .090\\ 
\hline
Recognized (Instr.)  & .85 & .108 & .72 & .109 & .79 & .109 \\  
\hline
\hline
Anthropomorphism & 2.69 & .631 & 3.11 & .632 & 2.51 & .632\\ 
\hline
Perceived Threat & 1.87 & .500 & 1.96 & .501 & 1.83 & .501 \\ 
\hline
Likability & 2.40 & .620 & 2.67 & .621 & 2.23 & .621 \\ 
\hline
\hline
Freq. RFR & .56 & .137 & .54 & .135 & .62 & .137 \\ 
\hline
Freq. CFR & .68 & .144 & .70 & .144 & .75 & .144\\
\hline
\hline
Avg $cRR$ & 15.75 & 5.863 & 11.07 & 5.863 & 12.95 & 5.863 \\ 
\hline
Avg $L$ & 5.53 & 2.312 & 5.17 & 2.311 & 5.74 & 2.312\\ 
\hline
Avg $L_{max}$ & 6.82 & 2.862 & 6.11 & 2.859 & 6.87 & 2.862\\ 
\hline
Avg $DET$  & 2.60 & 1.310 & 2.91 & 1.309 & 2.70 & 1.310 \\ 
\hline
\end{tabular}
\label{table:descriptive_humanlikeness}
\end{table*}

\subsection{Preliminary Analyses: Emotion Recognition}

\begin{figure*}[t!]
\centering
\includegraphics[width=.45\columnwidth]{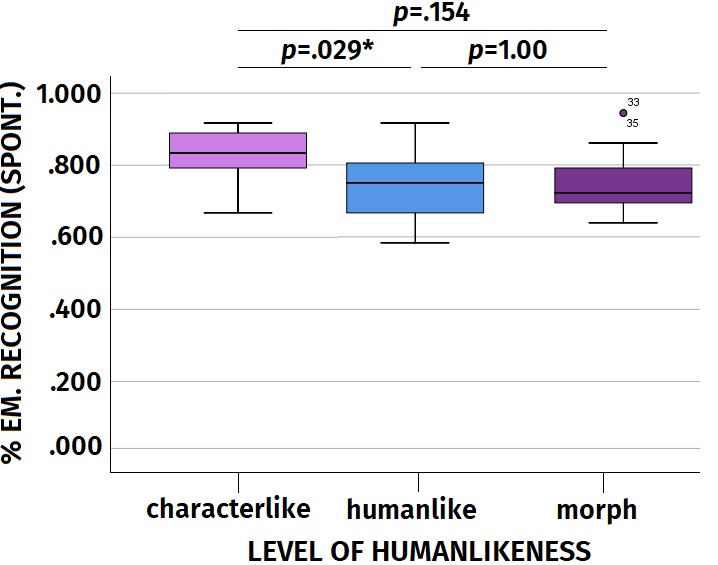}
\hspace{0.5 cm}
\includegraphics[width=.45\columnwidth]{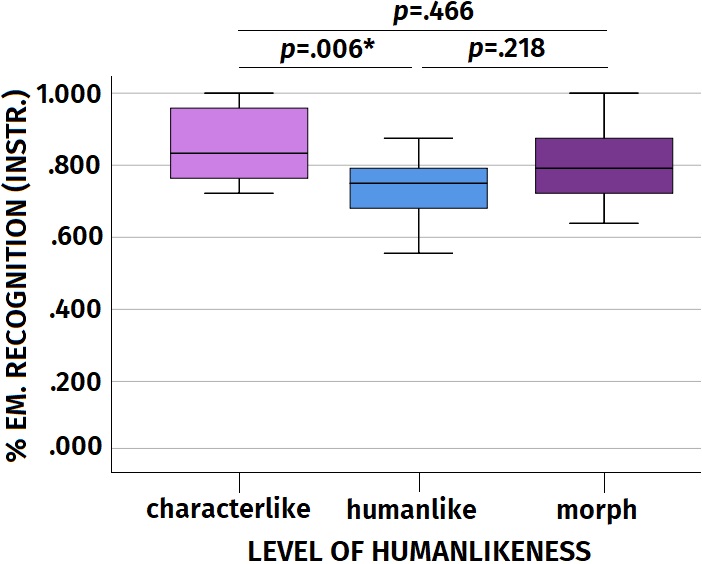}
\caption{Boxplots showing the effect of level of humanlikeness on the percentage of emotions correctly recognized for the spontaneous and instructed mimicry trials.}
\label{fig:EmotionRecognition_Spontaneous}
\end{figure*}

As a preliminary analysis, we checked whether participants' ability to recognize the emotions displayed by the artificial agents differed across embodiments and levels of humanlikeness. Interestingly, we discovered a main effect of humanlikess on the percentage of emotion recognized (\cf~Table \ref{table:3x3ANOVA} for the complete results). 
According to the results, participants' emotion recognition was better when participants observed the characterlike agents with respect to when they observed the humanlike agents (\cf~Table \ref{table:descriptive_humanlikeness} for the descriptive statistics). This was true both in the spontaneous mimicry ($p=.029$) and in the instructed mimicry trials ($p=.006$, \cf~Fig. \ref{fig:EmotionRecognition_Spontaneous}). No such differences in emotion recognition were observed between characterlike and morph agents (spontaneous mimicry trial: $p=.154$; instructed mimicry trial: $p=.466$) and between morph and humanlike agents across trials (spontaneous mimicry trial: $p=1.00$; instructed mimicry trial: $p=.218$).
When it comes to the 2x3 ANOVAs focusing on the agents' artificiality, we found a significant difference between artificial agents and the video-recorded human in terms of emotion recognition only for the instructed mimicry trial (\cf~Table \ref{table:ANOVA_descriptive_artificiality} for the results and descriptive statistics). In this case, the percentage of emotions correctly recognized was higher for the human with respect to the artificial agents.

\begin{table*}[]
\centering
\caption{\textbf{Results of 2x3 ANOVAs and Descriptive Statistics.} The significant results are displayed in bold. The Mean (\me) and standard deviation (\sd) of all dependent variables are divided per Artificial and Human Agents.}
\vspace{0.2cm}
\begin{tabular}{ | r || r | r | r || r | r || r | r |}
\hline
& \multicolumn{3}{c ||}{\textbf{Artificiality}} & \multicolumn{2}{c ||}{\textbf{Artif. Agents}} & \multicolumn{2}{c |}{\textbf{Human Video}}\\
\hline
\hline
\textbf{Agent's Percept.} & $F(1,41)$ & $p$ & $\eta p^2$ & \me & \sd & \me & \sd \\
\hline
Anthropomorphism &\textbf{ 130.064} & $<\textbf{.001}$ & \textbf{.760} & 2.77 & .659 & 4.13 & .761 \\ 
\hline
Perceived Threat & \textbf{29.800} & $<\textbf{.001}$ &\textbf{.421} & 1.93 & .544 & 1.51 & .451\\ 
\hline
Likability & \textbf{39.159} & $<\textbf{.001}$ & \textbf{.489} & 2.43 & .626 & 3.00 & .757 \\ 
\hline
\hline
\textbf{Emo. Recogn.}& $F(2, 40)$ & $p$ & $\eta p^2$ & \me & \sd & \me & \sd \\
\hline
Recognized (Spont.) & 1.470 & .233 & .035 & .78 & .096 & .75 & .151 \\ 
\hline
Recognized (Instr.) & \textbf{7.494} & \textbf{.009 }& \textbf{.158} & .79 & .119 & .84 & .120 \\
\hline
\hline
\textbf{Spont. Mimicry} & $F(1,32)$ & $p$ & $\eta p^2$ & \me & \sd & \me & \sd \\
\hline
Freq. RFR & \textbf{34.835} & $<\textbf{.001}$ & \textbf{.521} & .60 & .146 & .84 & .127 \\ 
\hline
Freq. CFR & \textbf{4.323} & \textbf{.046} & \textbf{.119 }& .74 & .144 & .79 & .156\\  
\hline
\hline
\textbf{Instr. Mimicry} & $F(1,39)$ & $p$ & $\eta p^2$ & \me & \sd & \me & \sd \\
\hline
Avg $cRR$ & .653 & .424 & .016 & 13.92 & 7.172 & 13.29 & 10.124 \\ 
\hline
Avg $L$ & .039 & .844 & .001 & 5.56 & 2.288 & 5.53  & 2.697 \\ 
\hline
Avg $L_{max}$ & .206 & .652 & .005 & 6.70 & 2.824 & 6.55 & 3.369 \\ 
\hline
Avg $DET$  & .249 & .621 & .006 & 2.75 & 1.270 & 2.65 & 1.638\\ 
\hline
\end{tabular}
\label{table:ANOVA_descriptive_artificiality}
\end{table*}

\vspace{0.2cm}\noindent\textbf{Discussion of Preliminary Analyses.} As predicted, the facial expressions of the video-recorded human were easier to recognize in comparison to the facial expressions of the artificial agents. However, somewhat unexpectedly, and partially in conflict with this result, the facial expressions of the characterlike artificial agents were easier to recognize with respect to those of the humanlike artificial agents both for the spontaneous and instructed mimicry trials. We ascribe this results to the stylized appearance of the characterlike agents, which might have made their expressions more readable and recognizable than those of the other agents.

\subsection{Results for Research Questions}

\subsubsection{Influence of Embodiment and Humanlikeness on Spontaneous Facial Mimicry [RQ1]}

Results disclosed a significant main effect of embodiment on spontaneous facial mimicry both for RFR and CFR (\cf~Table \ref{table:3x3ANOVA} for the complete results). However, we did not find any significant effect of humanlikeness and embodiment and humanlikeness alone on spontaneous facial mimicry.

\begin{figure*}[]
\centering
\includegraphics[width=.45\columnwidth]{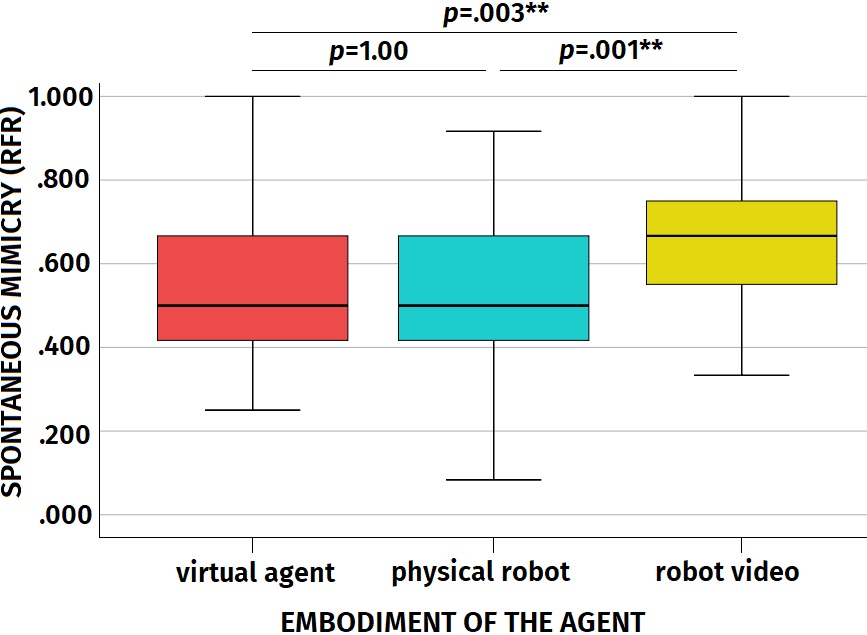}
\hspace{0.5 cm}
\includegraphics[width=.45\columnwidth]{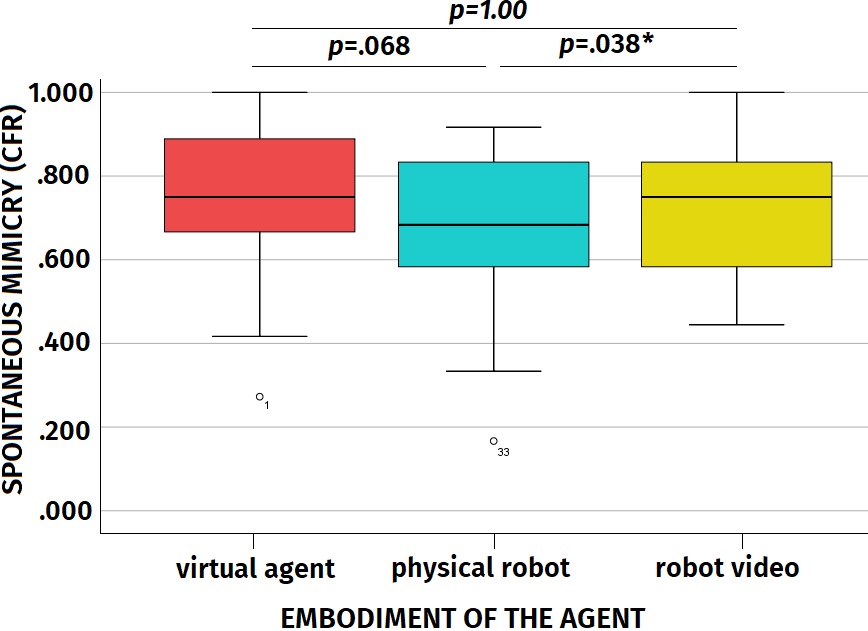}
\caption{Boxplots showing the effect of the agent's embodiment on frequency of spontaneous facial mimicry for RFR and CFR.}
\label{fig:SpontaneousMimicry_RFR}
\end{figure*}

Post-hoc analyses with a Bonferroni correction disclosed that for the RFR the video-recorded robot was mimicked significantly more than the virtual agent ($p=.003$, \cf~Table \ref{table:descriptive_embodiment} for the descriptive statistics and \cf~Fig. \ref{fig:SpontaneousMimicry_RFR} for the boxplot) and physical robot ($p=.001$), and that the physical robot and the virtual agent did not differ in spontaneous facial mimicry from each other ($p=1.00$). With regards to CFR, the post-hoc analyses showed that the physical robot was mimicked significantly less than the video-recorded robot ($p=.038$,  \cf~Fig. \ref{fig:SpontaneousMimicry_RFR} for the boxplot), while the virtual agent and the video-recorded robot did not differ in terms of spontaneous facial mimicry ($p=1.00$) and only a trend difference was present between the virtual agent and the physical robot ($p=.068$).

When taking into account the artificiality of the agent as the within-subject factor, we found a significant main effect of artificiality on spontaneous facial mimicry (\cf~Table \ref{table:ANOVA_descriptive_artificiality} for the results and the descriptive statistics). In this case, the video-recorded human was spontaneously mimicked significantly more than the artificial agent both for RFR and CFR.

\vspace{0.2cm}\noindent\textbf{Discussion of RQ1.} These results are somewhat complimentary to those we found for the manipulation checks. Indeed, it seems that the agent that elicited the highest ratings of co-presence, affective understanding, emotional interdependence, anthropomorphism, and likability, namely the physical robot, was also the agent that was spontaneously mimicked the least. If we take the facial mimicry-rapport hypothesis into account, this result is somewhat counterintuitive. Indeed, in line with this hypothesis, the robot eliciting the most favorable relational ratings should have been the one spontaneously mimicked the most. However, if we take the emotion recognition task into account, we can partially explain this result. Recognizing the emotions of another agent is an activity that implies putting some distance between the agent we observe and ourselves. It somewhat entails considering the agent we observe as a stimulus, rather than a relational agent. In this sense, we can hypothesize that an agent that is perceived as more socially present and elicits more positive perceptions is less good as a stimulus, it is more likely to act as a distractor, and hence can hinder the goal of the emotion recognition task. 
It is interesting to note that, when the agent is a human, this dynamic does not take place and the human is, as foreseeable, spontaneously mimicked more than the artificial agents. We can ascribe this result to the familiarity of the human stimulus. Indeed, the video-recorded human is undoubtedly more positively evaluated than the artificial agents and hence more likely to act as a distractor. However, it is also the stimulus with which participants are the most familiar and whose facial expressions they are more used to recognize.

\subsubsection{Influence of Embodiment and Humanlikeness on Instructed Facial Mimicry [RQ2]}

The results of the 3x3 repeated measures ANOVAs did not show any significant effect of embodiment and humanlikeness on instructed facial mimicry (\cf~Table \ref{table:3x3ANOVA} for the complete results). Similarly, the results of the 2x3 repeated measures ANOVA did not disclose any significant effect of the agents' artificiality on participants' instructed facial mimicry (\cf~Table \ref{table:ANOVA_descriptive_artificiality} for the results and the descriptive statistics). 

\vspace{0.2cm}\noindent\textbf{Discussion of RQ2.} As opposed to Hofree et al. \cite{hofree2014bridging}, in our study, the results of instructed facial mimicry are not congruent with those of spontaneous facial mimicry. Based on our findings, we can state that \textit{when facial mimicry is explicitly prompted, the agent's appearance and embodiment cease to have an influence on it}. This might be due to the fact that, when facial mimicry transforms itself into a purely imitative act, it loses its social value, and hence those variables that would have likely affected it due to their relational value, such as the agents' level of humanlikeness and their embodiment, do not influence it anymore. This assumption is further reinforced by the fact that people's ability to mimic an agent as closely as possible does not differ also when artificial and human agents are taken into account. 

\subsubsection{Spontaneous Facial Mimicry as Predictor of Perceived Social Presence and Perceptions of Artificial Agents [RQ3]}

The results of the regression analyses in Table \ref{table:regression_RQ3} show that spontaneous facial mimicry for RFR was a negative predictor of co-presence, attentional allocation, and affective understanding, whereas spontaneous facial mimicry for CFR was a negative predictor of attentional allocation and emotional interdependence. Moreover, they showed that spontaneous facial mimicry for RFR was a negative predictor of people's perceptions of the artificial agents' likability and anthropomorphism, and spontaneous facial mimicry for CFR was a negative predictor of participants' perception of the agents' likability (\cf~Table \ref{table:regression_RQ3} for the complete results). 

\begin{table*}[]
\centering
\caption{\textbf{Regression Analyses [RQ3].} Frequency of spontaneous facial mimicry as predictor of Social Presence and Perceptions of Artificial agents. The significant results are displayed in bold, while the trend effects are presented in italics.}
\vspace{0.2cm}
\begin{tabular}{ | r || r | r | r | r || r | r | r | r |}
\hline
& \multicolumn{4}{c ||}{\textbf{Spont. Mimicry (\% RFR)}} & \multicolumn{4}{c |}{\textbf{Spont. Mimicry (\% CFR)}} \\
Dependent Variables & $\beta$ & $t(149)$ &  $p$ & $r^2$ & $\beta$ & $t(154)$ & $p$ & $r^2$ \\
\hline
\hline
Co-presence &\textbf{ -.183} & \textbf{-2.271} &\textbf{ .025 }&\textbf{ .034} & -.100 & -1.246 & .215 & .010\\ 
\hline
Att. Allocation & \textbf{-.211} & \textbf{-2.626} &\textbf{ .010 }& \textbf{.045} & \textbf{-.230} & \textbf{-2.920} & \textbf{.004 }& \textbf{.053} \\ 
\hline
Aff. Understanding & \textbf{-.203} & \textbf{-2.527} &\textbf{.013} & \textbf{.041} & -.127 & -1.587 & .115 & .016 \\ 
\hline
Em. Interdependence & -.077 & -.940 & .349 & .006 & \textbf{-.204} & \textbf{2.581} &\textbf{ .011} & \textbf{.042}\\ 
\hline
Beha. Interdependence & -.076 & -.932 & .353 & .006 & -.098 & -1.224 & .223 & 010 \\
\hline
\hline
Anthropomorphism &\textbf{-.168} & \textbf{-2.074} &\textbf{.040} & \textbf{.028} & -.112 & -1.388 & .167 & .012\\ 
\hline
Perceived Threat & .055 & .666 & .506 & .003 & .124 & 1.546 & .124 & .015 \\ 
\hline
Likability &\textbf{-.191 }& \textbf{-2.373} &\textbf{.019} & \textbf{.037} & \textbf{-.262} & \textbf{-3.359} & \textbf{.001} & \textbf{.069} \\
\hline
\hline
\% Recognized & \textit{-.169} & \textit{-1.821} & \textit{.071} & \textit{.029} & -.073 & -.783 & .435 & .005 \\
\hline
Confidence Recogn. & \textbf{-.231} & \textbf{-2.520} & \textbf{.013 }& \textbf{.053} & -.121 & -1.310 & .193 & .015\\
\hline
\end{tabular}
\label{table:regression_RQ3}
\end{table*}

\vspace{0.2cm}\noindent\textbf{Discussion of RQ3.} These results are in line with those of RQ1 and seem to suggest that, in this study, \textit{the more the artificial agents were spontaneously mimicked, the less positive perceptions they elicited, the less socially co-present they were perceived, and the less people people felt emotionally connected with them and capable of understanding their affective states}. We assume that this result, which goes against most of the literature focusing on the social function of spontaneous facial mimicry, can be ascribed to the emotion recognition task in which participants were involved. Our hypothesis is that, within an emotion recognition task, spontaneous facial mimicry does not fulfill anymore a social function, but rather serves the purpose of emotion recognition. In this context, all the perceptual dimensions that are normally positively related to spontaneous facial mimicry become negatively related to mimicry as they act as distractors towards the ultimate emotion recognition goal of the task.

To verify this assumption, we performed a few additional regression analyses using spontaneous facial mimicry for RFR and CFR as predictors and the percentage of correctly recognized facial expressions and the confidence in the recognized emotion as dependent variables. As we supposed, participants' spontaneous facial mimicry was a significant negative predictor of their certainty of the correctness of the recognized emotion and a trend negative predictor of their emotion recognition performance (\cf~Table \ref{table:regression_RQ3} for the complete results). This indicates that \textit{the more participants spontaneously mimicked the artificial agents, the less they were confident in the emotion they recognized}. Such a result is particularly important as it corroborates the theory that facial mimicry serves the purpose of emotion recognition, but only when the emotions to recognize are ambiguous \cite{hess2001facial, fischer2012emotional}.

\subsubsection{Instructed Facial Mimicry as Predictor of Spontaneous Facial Mimicry [RQ4]}

The results of the regression analyses displayed in Table \ref{table:regression_RQ4} show that the average $cRR$, $L$, $L_{max}$, and $DET$ are all significant negative predictors of spontaneous facial mimicry for RFR but they do not equally predict spontaneous facial mimicry for CFR. 

\vspace{0.2cm}\noindent\textbf{Discussion RQ4} This result is extremely interesting as it suggests that, in this study, \textit{the more closely participants mimicked the facial expressions of the agents when instructed to do so, the less likely they were to spontaneously mimic the agents at an unconscious level of processing}.
Since we have seen that spontaneous facial mimicry for RFR was a negative predictor of participants' confidence in the recognized emotion (and partially also of their ability to recognize the target emotion), it does not surprise that people that mimic an emotion well under instruction, actually mimic it less at a subperceptual level. Indeed, if people are better able to mimic all the temporal dynamics of a target facial expression, they might also be more capable of recognizing that target emotion. In this sense, as opposed to spontaneous facial mimicry, instructed facial mimicry might signal a better understanding of the emotion. This finding entails that, even though in an emotion recognition task, instructed facial mimicry does not behave similarly to spontaneous facial mimicry, it still maintains a relation with it.

\begin{table*}[]
\centering
\caption{\textbf{Regression Analyses [RQ4].} Features of instructed facial mimicry as predictors of the frequency of Spontaneous facial mimicry. The significant results are displayed in bold.}
\vspace{0.2cm}
\begin{tabular}{ | r || r | r | r | r || r | r | r | r |}
\hline
& \multicolumn{4}{c ||}{\textbf{Spont. Mimicry (\% RFR)}} & \multicolumn{4}{c |}{\textbf{Spont. Mimicry (\% CFR)}} \\
Predictors & $\beta$ & $t(141)$ &  $p$ & $r^2$ & $\beta$ & $t(146)$ & $p$ & $r^2$ \\
\hline
\hline
Avg $cRR$ & \textbf{-.229 }& \textbf{-2.783} & \textbf{.006} & \textbf{.052} & -.093 & -1.124 & .263 & .009 \\ 
\hline
Avg $L$ & \textbf{-.186} & \textbf{-2.241 }& \textbf{.027} & \textbf{.035} & -.068 & -.819 & .414 & .005\\ 
\hline
Avg $L_{max}$ &\textbf{ -.175} &\textbf{ -2.107} &\textbf{ .037} & \textbf{.031} & -.057 & -.683 & .496 & .003 \\ 
\hline
Avg $DET$  & \textbf{-.212} & \textbf{-2.567} & \textbf{.011} & \textbf{.045} & -.098 & -1.189 & .236 & .010 \\ 
\hline
\end{tabular}
\label{table:regression_RQ4}
\end{table*}

\section{General Discussion}

This study investigated how the humanlikeness and embodiment of an artificial agent could influence people's mimicry of its facial expressions. Based on Hofree et al. \cite{hofree2014bridging}, we expected that physically embodied, co-present, and humanlike artificial agents could elicit higher spontaneous and instructed facial mimicry than virtually embodied, non-co-present, and less humanlike ones, and that instructed facial mimicry could positively predict spontaneous facial mimicry. Moreover, based on the link between facial mimicry and rapport, we postulated that spontaneous facial mimicry could positively predict participants' evaluations of the agents' anthropomorphism, social presence, and likability, and negatively predict their perceived uncanniness.  
Although our manipulation of embodiment and humanlikeness were successful, and the task we chose was taken from the existing literature \cite{hofree2014bridging, kulesza2015face}, the results we obtained did not meet our expectations (\cf~H1-H4 in section \ref{sec:RQs}). We found that: (i) the physically embodied, co-present artificial agent (i.e., the physical robot) was the one that was spontaneously mimicked the least regardless of its humanlikeness (\cf~H1); (ii) instructed facial mimicry did not behave congruently to spontaneous facial mimicry (\cf~H2); (iii) spontaneous facial mimicry negatively predicted anthropomorphism, social presence, and likability, and did not predict uncanniness (\cf~H3); and (iv) instructed facial mimicry negatively predicted spontaneous facial mimicry (\cf~H4). 

While these results were surprising, their consistency led to a hypothesis that some element of the task that was given to the participants hindered the social value of facial mimicry. Following the \textit{automatic embodiment account} \cite{niedenthal2010simulation}, we postulated that the task's focus on emotion recognition could have caused a change in the meaning of facial mimicry. Additional analyses confirmed our suspicion. In fact, they indicated that the spontaneous facial mimicry of the artificial agents was a significant negative predictor of participants' confidence in the emotion recognized. 
This result seems to suggest that, \textit{in the context of human-agent and human-robot mimicry, the emotion recognition goal of a task can flip the social value of spontaneous facial mimicry, and transform a physically embodied, co-present artificial agent into a distractor}. This may have arisen by chance due to elements of the study design and deserves further exploration and replication. 
The primary objective of this study was to understand whether spontaneous facial mimicry could be used as a cue of liking and rapport in HAI and HRI, and whether instructed facial mimicry could act as a proxy of spontaneous facial mimicry. 
Although our findings do not meet our expectations, the fact that they went in the exact opposite direction to our original hypotheses may suggest that, \textit{in an emotion recognition task, spontaneous facial mimicry can still be used as a predictor of liking and rapport, and instructed facial mimicry could still function as a predictor of spontaneous facial mimicry, but they need to be envisioned as negative predictors rather than positive ones}. Additional work is needed to corroborate these preliminary results, and understand whether context alone (emotion recognition task vs. social interaction) can influence the value of facial mimicry in HAI and HRI in the way we have described. 

\section{Limitations \& Future Work}

One limitation of the current experimental design is the focus on one particular robotic embodiment (i.e., the Furhat robot). While this platform has several advantages, like the easy alteration of facial features and expressions, it is sometimes difficult to discern facial detail clearly. 
By keeping the robot platform consistent across conditions, we could limit the influence of confounding factors on our results. However, this in turn reduced the strength of the manipulation of humanlikeness and could be the reason why we did not see the agents' anthropomorphism differ between the characterlike and humanlike, and the characterlike and morph conditions. 
Future studies should hence investigate facial mimicry in emotion recognition tasks carried out with multiple humanoid robots differing in their embodiment and degree of realism to check whether our findings still hold. We also suggest to replicate our study involving a larger and more diverse set of participants, particularly when it comes to academic background and gender.

Unlike most related work (e.g., \cite{hofree2014bridging}), in our experiment, we included stimuli covering all six basic emotions \cite{ekman1978facial}. For the analyses of facial mimicry, however, we combined people's responses to the different emotions together and calculated an average facial mimicry value. It is thus fair to assume that, while comprehensive, our results might not fit all six basic emotions equally. Another element of variation that one might need to control when studying facial mimicry is the observer's belief that an agent's facial expression reflects its subjective emotional state. 
In future facial mimicry studies, it would be interesting to include additional questionnaires capturing people's belief about the agent's emotional state when performing facial expressions, and their own emotional state before and after the experiment.

Since our study was task-based, non-interactive, and devoid on an emotional context, the acted nature of the agents' facial expressions was particularly clear. Future work should focus on bringing the study of facial mimicry into more interactive and social contexts and assess whether facial mimicry could be used in place of questionnaires to assess people's social attunement with artificial agents.  
An important pre-condition for using facial mimicry as a behavioral indicator of people's relationship with a robot is a robust and non-intrusive assessment technique of people's facial expressions. While the computer-vision-based approach discussed in this paper has shown promising results, further improvements are necessary to make it more robust with respect to different angles and light conditions. This is especially important if we want to bring the study of facial mimicry to less controlled scenarios.

\section{Conclusion}

In the study presented in this paper, we involved participants in an emotion recognition task carried out with artificial agents differing in their embodiment and degree of humanlikeness. In the first phase of the study, we asked participants to observe the artificial agents' facial expressions and attempt to identify the emotions they displayed. In the second phase of the study, instead, we asked participants to observe the agents' facial expressions, mimic them as closely as possible, and then identify them. We used the first part of the study to investigate the frequency of participants' spontaneous facial mimicry, and the second part to investigate the accuracy of their instructed facial mimicry. The aim was to understand whether spontaneous mimicry of artificial agents' facial expressions can be used as a behavioral cue of liking and rapport, and whether instructed facial mimicry could act as a proxy of its spontaneous counterpart. Our results suggest that, in an emotion recognition task, the physical instantiation of an artificial agent, together with its likability and anthropomorphism, intrudes rather than promotes people's spontaneous facial mimicry. Furthermore, results suggest that instructed facial mimicry negatively predicts spontaneous facial mimicry. Since the participants in this study mimicked the facial expressions of the artificial agents more when they were uncertain about the emotion to recognize, one possibility is that, in emotion recognition contexts, facial mimicry serves the purpose of emotion recognition. Even though our results did not support our initial hypotheses, they nevertheless show that spontaneous mimicry can be a behavioral cue of liking and rapport, and instructed facial mimicry a proxy of spontaneous facial mimicry.

\section*{Conflict of Interest Statement}
The authors declare that the research was conducted in the absence of any commercial or financial relationships that could be construed as a potential conflict of interest.

\section*{Author Contributions}

GP wrote the paper, formulated the research questions, segmented the stimuli for spontaneous mimicry, devised the methodology to analyze spontaneous mimicry, performed all the statistical analyses, created the tables and graphs, and interpreted the results. MPP collaborated in writing the paper, conceived and designed the study, wrote the program for the robot interaction, collected the data, segmented the stimuli for instructed mimicry, post-processed the spontaneous and instructed mimicry data, and helped with the interpretation of the results. IH conceived and designed the study, synthesized the facial expressions of the artificial agents, developed and ran the AU intensity detector, contributed to writing section 5, and helped with the processing of spontaneous and instructed mimicry.
GV conceived and designed the study, developed and ran the CRQA for the instructed mimicry, contributed to writing section 5, and helped with the processing of instructed mimicry.
MC, CP and GC gave advice on the design of the study, stimuli, read and gave comments on the final version of the paper.

\section*{Funding}
This work is partly supported by the Swedish Foundation for Strategic Research under the COIN project (RIT15-0133). The work received funding from ROMEO2 and Labex SMART (ANR-11-LABX-65) supported by French state funds managed by the ANR within the Investissements d’Avenir programme under reference ANR-11-IDEX-0004-02. The authors are solely responsible for the content of this publication. 

\section*{Acknowledgments}
Thanks to A. Axelsson, R. Cort, A. Y. Gao, D. Golay, K. von Hausswolff, A. Jansson, M. Lind, I. L\"oscher, G. Nauwerck and A. Persson for input on the experimental design of the experiment. I. Hupont partly undertook this work before arriving to European Commission's Joint Research Centre (JRC, Seville), but finalised in the JRC and in line with the JRC work program. She participated in the experimental part of the study when working at Institut des Syst\`emes Intelligents et de Robotique (Sorbonne Universit\'e, Paris) and in the writing of the paper after joining JRC.

\section*{Data Availability Statement}
The dataset presented in this article is not readily available because it contains personal information on the participants. An anonymized version of the data supporting the conclusions of this article can be made available by the authors upon request.

\bibliography{refs}

\begin{thebibliography}{76}
\providecommand{\natexlab}[1]{#1}
\providecommand{\url}[1]{\texttt{#1}}
\expandafter\ifx\csname urlstyle\endcsname\relax
  \providecommand{\doi}[1]{doi: #1}\else
  \providecommand{\doi}{doi: \begingroup \urlstyle{rm}\Url}\fi

\bibitem[Aifanti et~al.(2010)Aifanti, Papachristou, and
  Delopoulos]{aifanti2010mug}
Niki Aifanti, Christos Papachristou, and Anastasios Delopoulos.
\newblock {The MUG facial expression database}.
\newblock In \emph{IEEE Int. Workshop on Image Analysis for Multimedia
  Interactive Services}, pages 1--4, 2010.

\bibitem[Al~Moubayed et~al.(2012)Al~Moubayed, Beskow, Skantze, and
  Granstr{\"o}m]{almoubayed2012furhat}
Samer Al~Moubayed, Jonas Beskow, Gabriel Skantze, and Bj{\"o}rn Granstr{\"o}m.
\newblock {Furhat: a back-projected human-like robot head for multiparty
  human-machine interaction}.
\newblock In \emph{Cognitive Behavioural Systems}, pages 114--130. 2012.

\bibitem[Bartneck et~al.(2009)Bartneck, Kuli{\'c}, Croft, and
  Zoghbi]{bartneck2009measurement}
Christoph Bartneck, Dana Kuli{\'c}, Elizabeth Croft, and Susana Zoghbi.
\newblock Measurement instruments for the anthropomorphism, animacy,
  likeability, perceived intelligence, and perceived safety of robots.
\newblock \emph{Int. Journal of Social Robotics}, 1\penalty0 (1):\penalty0
  71--81, 2009.

\bibitem[Bavelas et~al.(1986)Bavelas, Black, Lemery, and
  Mullett]{bavelas1986show}
Janet~B Bavelas, Alex Black, Charles~R Lemery, and Jennifer Mullett.
\newblock " i show how you feel": Motor mimicry as a communicative act.
\newblock \emph{Journal of personality and social psychology}, 50\penalty0
  (2):\penalty0 322, 1986.

\bibitem[Bernieri(1988)]{bernieri1988coordinated}
Frank~J Bernieri.
\newblock Coordinated movement and rapport in teacher-student interactions.
\newblock \emph{Journal of Nonverbal behavior}, 12\penalty0 (2):\penalty0
  120--138, 1988.

\bibitem[Bourgeois and Hess(2008)]{bourgeois2008impact}
Patrick Bourgeois and Ursula Hess.
\newblock The impact of social context on mimicry.
\newblock \emph{Biological Psychology}, 77\penalty0 (3):\penalty0 343--352,
  2008.

\bibitem[Calvo-Barajas et~al.(2020)Calvo-Barajas, Perugia, and
  Castellano]{calvo2020effects}
Natalia Calvo-Barajas, Giulia Perugia, and Ginevra Castellano.
\newblock The effects of robot’s facial expressions on children’s first
  impressions of trustworthiness.
\newblock In \emph{2020 29th IEEE International Conference on Robot and Human
  Interactive Communication (RO-MAN)}, pages 165--171. IEEE, 2020.

\bibitem[Chartrand and Bargh(1999)]{chartrand1999chameleon}
Tanya~L Chartrand and John~A Bargh.
\newblock The chameleon effect: the perception--behavior link and social
  interaction.
\newblock \emph{Journal of personality and social psychology}, 76\penalty0
  (6):\penalty0 893, 1999.

\bibitem[Chung and Cakmak(2018)]{chung2018your}
Michael Jae-Yoon Chung and Maya Cakmak.
\newblock “how was your stay?”: Exploring the use of robots for gathering
  customer feedback in the hospitality industry.
\newblock In \emph{2018 27th IEEE International Symposium on Robot and Human
  Interactive Communication (RO-MAN)}, pages 947--954. IEEE, 2018.

\bibitem[Dalal and Triggs(2005)]{HOG}
Navneet Dalal and Bill Triggs.
\newblock Histograms of oriented gradients for human detection.
\newblock In \emph{IEEE Computer Society Conference on Computer Vision and
  Pattern Recognition}, volume~1, pages 886--893, 2005.

\bibitem[Darwin and Prodger(1998)]{darwin1998expression}
Charles Darwin and Phillip Prodger.
\newblock \emph{The expression of the emotions in man and animals}.
\newblock Oxford University Press, USA, 1998.

\bibitem[Davis et~al.(1980)]{davis1980multidimensional}
Mark~H Davis et~al.
\newblock A multidimensional approach to individual differences in empathy.
\newblock \emph{JSAS Catalog of Selected Documents in Psychology}, 1980.

\bibitem[Dimberg(1990)]{dimberg1990distinguished}
Ulf Dimberg.
\newblock For distinguished early career contribution to psychophysiology:
  award address, 1988: facial electromyography and emotional reactions.
\newblock \emph{psychophysiology}, 27\penalty0 (5):\penalty0 481--494, 1990.

\bibitem[Dimberg(1997)]{dimberg1997facial}
Ulf Dimberg.
\newblock Facial reactions: Rapidly evoked emotional responses.
\newblock \emph{Journal of Psychophysiology}, 1997.

\bibitem[Dimberg and Thunberg(1998)]{dimberg1998rapid}
Ulf Dimberg and Monika Thunberg.
\newblock Rapid facial reactions to emotional facial expressions.
\newblock \emph{Scandinavian journal of psychology}, 39\penalty0 (1):\penalty0
  39--45, 1998.

\bibitem[Dimberg et~al.(2000)Dimberg, Thunberg, and
  Elmehed]{dimberg2000unconscious}
Ulf Dimberg, Monika Thunberg, and Kurt Elmehed.
\newblock Unconscious facial reactions to emotional facial expressions.
\newblock \emph{Psychological science}, 11\penalty0 (1):\penalty0 86--89, 2000.

\bibitem[Ekman and Rosenberg(1997)]{ekman1997face}
Paul Ekman and Erika~L Rosenberg.
\newblock \emph{What the face reveals: Basic and applied studies of spontaneous
  expression using the Facial Action Coding System (FACS)}.
\newblock Oxford University Press, USA, 1997.

\bibitem[Ekman et~al.(1978)Ekman, Friesen, and Hager]{ekman1978facial}
Paul Ekman, Wallace~V Friesen, and Joseph~C Hager.
\newblock \emph{Facial action coding system: Investigator's guide}.
\newblock Consulting Psychologists Press, 1978.

\bibitem[Ekman et~al.(1983)Ekman, Levenson, and Friesen]{ekman1983autonomic}
Paul Ekman, Robert~W Levenson, and Wallace~V Friesen.
\newblock Autonomic nervous system activity distinguishes among emotions.
\newblock \emph{science}, 221\penalty0 (4616):\penalty0 1208--1210, 1983.

\bibitem[Fasel and Luettin(2003)]{FASEL2003259}
B.~Fasel and Juergen Luettin.
\newblock Automatic facial expression analysis: a survey.
\newblock \emph{Pattern Recognition}, 36\penalty0 (1):\penalty0 259--275, 2003.

\bibitem[Fischer et~al.(2012)Fischer, Becker, and
  Veenstra]{fischer2012emotional}
Agneta Fischer, Daniela Becker, and Lotte Veenstra.
\newblock Emotional mimicry in social context: the case of disgust and pride.
\newblock \emph{Frontiers in psychology}, 3:\penalty0 475, 2012.

\bibitem[Gratch et~al.(2006)Gratch, Okhmatovskaia, Lamothe, Marsella, Morales,
  van~der Werf, and Morency]{gratch2006virtual}
Jonathan Gratch, Anna Okhmatovskaia, Francois Lamothe, Stacy Marsella, Mathieu
  Morales, Rick~J van~der Werf, and Louis-Philippe Morency.
\newblock Virtual rapport.
\newblock In \emph{Int. Workshop on Intelligent Virtual Agents}, pages 14--27,
  2006.

\bibitem[Hager et~al.(2002)Hager, Ekman, and Friesen]{EF02}
Joseph~C Hager, Paul Ekman, and Wallace~V Friesen.
\newblock Facial {A}ction {C}oding {S}ystem.
\newblock \emph{Salt Lake City, UT: A Human Face}, 2002.

\bibitem[Hall(1969)]{hall_hidden_1969}
E.~T. Hall.
\newblock \emph{{The Hidden Dimension}}, volume 1990.
\newblock Anchor Books New York, 1969.

\bibitem[Harms and Biocca(2004)]{harms2004internal}
Chad Harms and Frank Biocca.
\newblock Internal consistency and reliability of the networked minds measure
  of social presence.
\newblock 2004.

\bibitem[Hatfield et~al.(1992)Hatfield, Cacioppo, and
  Rapson]{hatfield1992primitive}
Elaine Hatfield, John~T Cacioppo, and Richard~L Rapson.
\newblock Primitive emotional contagion.
\newblock \emph{Review of personality and social psychology}, 14:\penalty0
  151--177, 1992.

\bibitem[Hatfield et~al.(1993)Hatfield, Cacioppo, and
  Rapson]{hatfield1993emotional}
Elaine Hatfield, John~T Cacioppo, and Richard~L Rapson.
\newblock Emotional contagion.
\newblock \emph{Current directions in psychological science}, 2\penalty0
  (3):\penalty0 96--100, 1993.

\bibitem[Hawk et~al.(2012)Hawk, Fischer, and Van~Kleef]{hawk2012face}
Skyler~T Hawk, Agneta~H Fischer, and Gerben~A Van~Kleef.
\newblock Face the noise: Embodied responses to nonverbal vocalizations of
  discrete emotions.
\newblock \emph{Journal of personality and social psychology}, 102\penalty0
  (4):\penalty0 796, 2012.

\bibitem[Hess and Blairy(2001)]{hess2001facial}
Ursula Hess and Sylvie Blairy.
\newblock Facial mimicry and emotional contagion to dynamic emotional facial
  expressions and their influence on decoding accuracy.
\newblock \emph{International journal of psychophysiology}, 40\penalty0
  (2):\penalty0 129--141, 2001.

\bibitem[Hess and Fischer(2013)]{hess2013}
Ursula Hess and Agneta Fischer.
\newblock Emotional mimicry as social regulation.
\newblock \emph{Personality and Social Psychology Review}, 17\penalty0
  (2):\penalty0 142--157, 2013.

\bibitem[Hess and Fischer(2014)]{hess2014}
Ursula Hess and Agneta Fischer.
\newblock Emotional mimicry: Why and when we mimic emotions.
\newblock \emph{Social and Personality Psychology Compass}, 8\penalty0
  (2):\penalty0 45--57, 2014.

\bibitem[Hess et~al.(1995)Hess, Banse, and Kappas]{hess1995intensity}
Ursula Hess, Rainer Banse, and Arvid Kappas.
\newblock The intensity of facial expression is determined by underlying
  affective state and social situation.
\newblock \emph{Journal of personality and social psychology}, 69\penalty0
  (2):\penalty0 280, 1995.

\bibitem[Heyes(2011)]{heyes2011automatic}
Cecilia Heyes.
\newblock Automatic imitation.
\newblock \emph{Psychological bulletin}, 137\penalty0 (3):\penalty0 463, 2011.

\bibitem[Ho and MacDorman(2010)]{ho2010revisiting}
Chin-Chang Ho and Karl~F MacDorman.
\newblock Revisiting the uncanny valley theory: Developing and validating an
  alternative to the godspeed indices.
\newblock \emph{Computers in Human Behavior}, 26\penalty0 (6):\penalty0
  1508--1518, 2010.

\bibitem[Hoegen et~al.(2018)Hoegen, Van Der~Schalk, Lucas, and
  Gratch]{hoegen2018impact}
Rens Hoegen, Job Van Der~Schalk, Gale Lucas, and Jonathan Gratch.
\newblock The impact of agent facial mimicry on social behavior in a prisoner's
  dilemma.
\newblock In \emph{Proceedings of the 18th International Conference on
  Intelligent Virtual Agents}, pages 275--280, 2018.

\bibitem[Hofree et~al.(2014)Hofree, Ruvolo, Bartlett, and
  Winkielman]{hofree2014bridging}
Galit Hofree, Paul Ruvolo, Marian~Stewart Bartlett, and Piotr Winkielman.
\newblock Bridging the mechanical and the human mind: spontaneous mimicry of a
  physically present android.
\newblock \emph{PloS one}, 9\penalty0 (7), 2014.

\bibitem[Hofree et~al.(2018)Hofree, Ruvolo, Reinert, Bartlett, and
  Winkielman]{hofree2018behind}
Galit Hofree, Paul Ruvolo, Audrey Reinert, Marian~S Bartlett, and Piotr
  Winkielman.
\newblock Behind the robot’s smiles and frowns: In social context, people do
  not mirror android’s expressions but react to their informational value.
\newblock \emph{Frontiers in neurorobotics}, 12:\penalty0 14, 2018.

\bibitem[Hupont and Chetouani(2019)]{hupont2019}
Isabelle Hupont and Mohamed Chetouani.
\newblock Region-based facial representation for real-time action units
  intensity detection across datasets.
\newblock \emph{Pattern Analysis and Applications}, 22\penalty0 (2):\penalty0
  477--489, 2019.

\bibitem[Ito et~al.(2004)Ito, Murano, and Gomi]{ito2004fast}
Takayuki Ito, Emi~Z Murano, and Hiroaki Gomi.
\newblock Fast force-generation dynamics of human articulatory muscles.
\newblock \emph{Journal of applied physiology}, 96\penalty0 (6):\penalty0
  2318--2324, 2004.

\bibitem[Izard(2013)]{izard2013human}
Carroll~E Izard.
\newblock \emph{Human emotions}.
\newblock Springer Science \& Business Media, 2013.

\bibitem[Jaques et~al.(2016)Jaques, McDuff, Kim, and
  Picard]{jaques2016understanding}
Natasha Jaques, Daniel McDuff, Yoo~Lim Kim, and Rosalind Picard.
\newblock Understanding and predicting bonding in conversations using thin
  slices of facial expressions and body language.
\newblock In \emph{International Conference on Intelligent Virtual Agents},
  pages 64--74. Springer, 2016.

\bibitem[K{\"a}tsyri et~al.(2015)K{\"a}tsyri, F{\"o}rger,
  M{\"a}k{\"a}r{\"a}inen, and Takala]{katsyri2015review}
Jari K{\"a}tsyri, Klaus F{\"o}rger, Meeri M{\"a}k{\"a}r{\"a}inen, and Tapio
  Takala.
\newblock A review of empirical evidence on different uncanny valley
  hypotheses: support for perceptual mismatch as one road to the valley of
  eeriness.
\newblock \emph{Frontiers in Psychology}, 6, 2015.

\bibitem[Kulesza et~al.(2015)Kulesza, Cis{\l}ak, Vallacher, Nowak, Czekiel, and
  Bedynska]{kulesza2015face}
Wojciech~Marek Kulesza, Aleksandra Cis{\l}ak, Robin~R Vallacher, Andrzej Nowak,
  Martyna Czekiel, and Sylwia Bedynska.
\newblock The face of the chameleon: The experience of facial mimicry for the
  mimicker and the mimickee.
\newblock \emph{The Journal of Social Psychology}, 155\penalty0 (6):\penalty0
  590--604, 2015.

\bibitem[Laird and Bresler(1992)]{laird1992process}
James~D Laird and Charles Bresler.
\newblock The process of emotional experience: A self-perception theory.
\newblock 1992.

\bibitem[Lakin et~al.(2003)Lakin, Jefferis, Cheng, and
  Chartrand]{lakin2003chameleon}
Jessica~L Lakin, Valerie~E Jefferis, Clara~Michelle Cheng, and Tanya~L
  Chartrand.
\newblock The chameleon effect as social glue: Evidence for the evolutionary
  significance of nonconscious mimicry.
\newblock \emph{Journal of Nonverbal Behavior}, 27\penalty0 (3):\penalty0
  145--162, 2003.

\bibitem[Lanzetta and Englis(1989)]{lanzetta1989expectations}
John~T Lanzetta and Basil~G Englis.
\newblock Expectations of cooperation and competition and their effects on
  observers' vicarious emotional responses.
\newblock \emph{Journal of personality and social psychology}, 56\penalty0
  (4):\penalty0 543, 1989.

\bibitem[Li(2015)]{li2015benefit}
Jamy Li.
\newblock The benefit of being physically present: A survey of experimental
  works comparing copresent robots, telepresent robots and virtual agents.
\newblock \emph{Int. Journal of Human-Computer Studies}, 77:\penalty0 23--37,
  2015.

\bibitem[Marwan et~al.(2007)Marwan, Romano, Thiel, and Kurths]{Marwan07}
Norbert Marwan, M.~Carmen Romano, Marco Thiel, and Jürgen Kurths.
\newblock Recurrence plots for the analysis of complex systems.
\newblock \emph{Physics Reports}, 438\penalty0 (5–6):\penalty0 237 -- 329,
  2007.

\bibitem[Mattheij et~al.(2013)Mattheij, Nilsenova, and
  Postma]{mattheij2013vocal}
Ruud Mattheij, Marie Nilsenova, and Eric Postma.
\newblock Vocal and facial imitation of humans interacting with virtual agents.
\newblock In \emph{2013 Humaine Association Conference on Affective Computing
  and Intelligent Interaction}, pages 815--820. IEEE, 2013.

\bibitem[Mattheij et~al.(2015)Mattheij, Postma-Nilsenov{\'a}, and
  Postma]{mattheij2015mirror}
Ruud Mattheij, Marie Postma-Nilsenov{\'a}, and Eric Postma.
\newblock Mirror mirror on the wall.
\newblock \emph{Journal of Ambient Intelligence and Smart Environments},
  7\penalty0 (2):\penalty0 121--132, 2015.

\bibitem[Mavadati et~al.(2013)Mavadati, Mahoor, Bartlett, Trinh, and
  Cohn]{DISFA}
S~Mohammad Mavadati, Mohammad~H Mahoor, Kevin Bartlett, Philip Trinh, and
  Jeffrey~F Cohn.
\newblock {DISFA}: A spontaneous facial action intensity database.
\newblock \emph{IEEE Transactions on Affective Computing}, 4\penalty0
  (2):\penalty0 151--160, 2013.

\bibitem[McIntosh et~al.(2006)McIntosh, Reichmann-Decker, Winkielman, and
  Wilbarger]{mcintosh2006social}
Daniel~N McIntosh, Aimee Reichmann-Decker, Piotr Winkielman, and Julia~L
  Wilbarger.
\newblock When the social mirror breaks: deficits in automatic, but not
  voluntary, mimicry of emotional facial expressions in autism.
\newblock \emph{Developmental science}, 9\penalty0 (3):\penalty0 295--302,
  2006.

\bibitem[Moody et~al.(2007)Moody, McIntosh, Mann, and Weisser]{moody2007more}
Eric~J Moody, Daniel~N McIntosh, Laura~J Mann, and Kimberly~R Weisser.
\newblock More than mere mimicry? the influence of emotion on rapid facial
  reactions to faces.
\newblock \emph{Emotion}, 7\penalty0 (2):\penalty0 447, 2007.

\bibitem[Mori et~al.(2012)Mori, MacDorman, and Kageki]{mori2012uncanny}
Masahiro Mori, Karl~F MacDorman, and Norri Kageki.
\newblock The uncanny valley [from the field].
\newblock \emph{IEEE Robotics \& Automation Magazine}, 19\penalty0
  (2):\penalty0 98--100, 2012.

\bibitem[Niedenthal et~al.(2001)Niedenthal, Brauer, Halberstadt, and
  Innes-Ker]{niedenthal2001did}
Paula~M Niedenthal, Markus Brauer, Jamin~B Halberstadt, and {\AA}se~H
  Innes-Ker.
\newblock When did her smile drop? facial mimicry and the influences of
  emotional state on the detection of change in emotional expression.
\newblock \emph{Cognition \& Emotion}, 15\penalty0 (6):\penalty0 853--864,
  2001.

\bibitem[Niedenthal et~al.(2010)Niedenthal, Mermillod, Maringer, and
  Hess]{niedenthal2010simulation}
Paula~M Niedenthal, Martial Mermillod, Marcus Maringer, and Ursula Hess.
\newblock The simulation of smiles (sims) model: Embodied simulation and the
  meaning of facial expression.
\newblock \emph{Behavioral and brain sciences}, 33\penalty0 (6):\penalty0 417,
  2010.

\bibitem[Numata et~al.(2020)Numata, Sato, Asa, Koike, Miyata, Nakagawa, Sumiya,
  and Sadato]{numata2020achieving}
Takashi Numata, Hiroki Sato, Yasuhiro Asa, Takahiko Koike, Kohei Miyata, Eri
  Nakagawa, Motofumi Sumiya, and Norihiro Sadato.
\newblock Achieving affective human--virtual agent communication by enabling
  virtual agents to imitate positive expressions.
\newblock \emph{Scientific reports}, 10\penalty0 (1):\penalty0 1--11, 2020.

\bibitem[Paetzel et~al.(2016)Paetzel, Peters, Nystr\"{o}m, and
  Castellano]{paetzel:icsr}
Maike Paetzel, Christopher Peters, Ingela Nystr\"{o}m, and Ginevra Castellano.
\newblock {Congruency matters - How ambiguous gender cues increase a robot's
  uncanniness}.
\newblock In \emph{International Conference on Social Robotics}, 2016.

\bibitem[Paetzel et~al.(2017)Paetzel, Varni, Hupont, Chetouani, Peters, and
  Castellano]{paetzel2017investigating}
Maike Paetzel, Giovanna Varni, Isabelle Hupont, Mohamed Chetouani, Christopher
  Peters, and Ginevra Castellano.
\newblock Investigating the influence of embodiment on facial mimicry in hri
  using computer vision-based measures.
\newblock In \emph{2017 26th IEEE International Symposium on Robot and Human
  Interactive Communication (RO-MAN)}, pages 579--586. IEEE, 2017.

\bibitem[Paetzel et~al.(2020)Paetzel, Perugia, and
  Castellano]{paetzel2020persistence}
Maike Paetzel, Giulia Perugia, and Ginevra Castellano.
\newblock The persistence of first impressions: The effect of repeated
  interactions on the perception of a social robot.
\newblock In \emph{Proceedings of the 2020 ACM/IEEE International Conference on
  Human-Robot Interaction}, pages 73--82, 2020.

\bibitem[Paetzel-Pr{\"u}smann et~al.(2021)Paetzel-Pr{\"u}smann, Perugia, and
  Castellano]{paetzel2021influence}
Maike Paetzel-Pr{\"u}smann, Giulia Perugia, and Ginevra Castellano.
\newblock The influence of robot personality on the development of uncanny
  feelings.
\newblock \emph{Computers in Human Behavior}, 120:\penalty0 106756, 2021.

\bibitem[Perugia et~al.(2020{\natexlab{a}})Perugia, D{\'\i}az-Boladeras,
  Catal{\`a}-Mallofr{\'e}, Barakova, and Rauterberg]{perugia2020engage}
Giulia Perugia, Marta D{\'\i}az-Boladeras, Andreu Catal{\`a}-Mallofr{\'e},
  Emilia~I Barakova, and Matthias Rauterberg.
\newblock Engage-dem: a model of engagement of people with dementia.
\newblock \emph{IEEE Transactions on Affective Computing}, 2020{\natexlab{a}}.

\bibitem[Perugia et~al.(2020{\natexlab{b}})Perugia, Paetzel, and
  Castellano]{perugia2020role}
Giulia Perugia, Maike Paetzel, and Ginevra Castellano.
\newblock On the role of personality and empathy in human-human, human-agent,
  and human-robot mimicry.
\newblock In \emph{International Conference on Social Robotics}, pages
  120--131. Springer, 2020{\natexlab{b}}.

\bibitem[Perugia et~al.(2021)Perugia, Paetzel-Prüsmann, Alanenpää, and
  Castellano]{perugia2021ICanSeeIt}
Giulia Perugia, Maike Paetzel-Prüsmann, Madelene Alanenpää, and Ginevra
  Castellano.
\newblock I can see it in your eyes: Gaze as an implicit cue of uncanniness and
  task performance in repeated interactions with robots.
\newblock \emph{Frontiers in Robotics and AI}, 8:\penalty0 78, 2021.

\bibitem[Philip et~al.(2018)Philip, Martin, and Clavel]{philip2018rapid}
Leonor Philip, Jean-Claude Martin, and C{\'e}line Clavel.
\newblock Rapid facial reactions in response to facial expressions of emotion
  displayed by real versus virtual faces.
\newblock \emph{i-Perception}, 9\penalty0 (4):\penalty0 2041669518786527, 2018.

\bibitem[Prepin et~al.(2013)Prepin, Ochs, and Pelachaud]{prepin2013beyond}
Ken Prepin, Magalie Ochs, and Catherine Pelachaud.
\newblock Beyond backchannels: co-construction of dyadic stancce by reciprocal
  reinforcement of smiles between virtual agents.
\newblock In \emph{Proceedings of the Annual Meeting of the Cognitive Science
  Society}, volume~35, 2013.

\bibitem[Rammstedt and John(2007)]{rammstedt2007measuring}
Beatrice Rammstedt and Oliver~P John.
\newblock Measuring personality in one minute or less: A 10-item short version
  of the big five inventory in english and german.
\newblock \emph{Journal of Research in Personality}, 41\penalty0 (1):\penalty0
  203--212, 2007.

\bibitem[Riek et~al.(2009)Riek, Rabinowitch, Chakrabarti, and
  Robinson]{riek2009anthropomorphism}
Laurel~D Riek, Tal-Chen Rabinowitch, Bhismadev Chakrabarti, and Peter Robinson.
\newblock How anthropomorphism affects empathy toward robots.
\newblock In \emph{ACM/IEEE Int. Conference on Human Robot Interaction}, pages
  245--246, 2009.

\bibitem[Rosenthal-von~der P{\"u}tten and
  Kr{\"a}mer(2014)]{rosenthal2014design}
Astrid~M Rosenthal-von~der P{\"u}tten and Nicole~C Kr{\"a}mer.
\newblock How design characteristics of robots determine evaluation and uncanny
  valley related responses.
\newblock \emph{Computers in Human Behavior}, 36:\penalty0 422--439, 2014.

\bibitem[Skantze and Al~Moubayed(2012)]{skantze2012iristk}
Gabriel Skantze and Samer Al~Moubayed.
\newblock {IrisTK: a Statechart-based Toolkit for Multi-party Face-to-face
  Interaction}.
\newblock In \emph{Int. Conference on Multimodal Interaction}, pages 69--76,
  2012.

\bibitem[Tickle-Degnen and Rosenthal(1990)]{tickle1990nature}
Linda Tickle-Degnen and Robert Rosenthal.
\newblock The nature of rapport and its nonverbal correlates.
\newblock \emph{Psychological inquiry}, 1\penalty0 (4):\penalty0 285--293,
  1990.

\bibitem[Tomkins(1984)]{tomkins1984affect}
Silvan~S Tomkins.
\newblock Affect theory.
\newblock \emph{Approaches to emotion}, 163\penalty0 (163-195), 1984.

\bibitem[Varni et~al.(2017)Varni, Hupont, Clavel, and
  Chetouani]{varni2017computational}
Giovanna Varni, Isabelle Hupont, Chloe Clavel, and Mohamed Chetouani.
\newblock Computational study of primitive emotional contagion in dyadic
  interactions.
\newblock \emph{IEEE Transactions on Affective Computing}, 11\penalty0
  (2):\penalty0 258--271, 2017.

\bibitem[Viola and Jones(2004)]{VJ04}
Paul Viola and Michael~J Jones.
\newblock Robust real-time face detection.
\newblock \emph{International Journal of Computer Vision}, 57\penalty0
  (2):\penalty0 137--154, 2004.

\bibitem[Wang and Gratch(2009)]{wang2009rapport}
Ning Wang and Jonathan Gratch.
\newblock Rapport and facial expression.
\newblock In \emph{2009 3rd International Conference on Affective Computing and
  Intelligent Interaction and Workshops}, pages 1--6. IEEE, 2009.

\bibitem[Xiong and Torre(2013)]{XD13}
Xuehan Xiong and Fernando Torre.
\newblock Supervised descent method and its applications to face alignment.
\newblock In \emph{IEEE Conference on Computer Vision and Pattern Recognition},
  pages 532--539, 2013.

\end{thebibliography}

\end{document}